\documentclass{optica-article}
\journal{opticajournal} 
\articletype{Research Article}

\usepackage{lineno}
\usepackage{bm}
\newcommand{\LL}{\mathrm{L}}
\newcommand{\dd}{\mathrm{d}}
\newcommand{\s}{\mathrm{s}}
\newcommand{\B}{\mathrm{B}}

\begin{document}

\title{Static-to-dynamic field conversion in a temporally switched Lorentz medium}

\author{Alazar G. Salfore and Mario J. Mencagli}

\address{Department of Electrical and Computer Engineering, University of Delaware, Newark, DE 19716, USA}

\email{mencagli@udel.edu}


\begin{abstract*} 
We investigate static-to-dynamic electromagnetic field conversion in a temporally switched Lorentz medium initially subjected to an electrostatic field. An abrupt change of the oscillator strength drives the material away from its initial equilibrium, while the polarization retains memory of the pre-switch state and acts as the source of the resulting transient. We formulate the resulting initial-value problem in the Laplace domain and derive an analytical representation of the generated field in terms of the poles of the finite Lorentz slab. Material dispersion gives rise to a richer modal structure than in the nondispersive case, including families of slab-mode poles that accumulate toward the singularities associated with the Lorentz polarization dynamics. We further show that material memory governs the earliest stage of the transient, producing a smooth field onset and setting the velocity of the earliest propagating disturbance through the high-frequency permittivity. Because the temporal switch generates broadband spectral content, the transient samples positive-, near-zero-, and negative-permittivity regions of the post-switch material response, leading to qualitatively different spatial field distributions. These results reveal how material dispersion and memory fundamentally shape the generation and evolution of radiation produced from an initially static electromagnetic state.
\end{abstract*}

\section{Introduction}
Time-varying electromagnetic media have been investigated for decades as a means of controlling waves through externally driven changes of their constitutive properties. Early studies of temporally varying dielectrics and plasmas established the basic principles governing wave propagation and scattering in nonstationary media \cite{Morgenthaler58,Felsen70,Fante71,Oliner61}. More recently, advances in metamaterials, metasurfaces, and rapidly tunable platforms have enabled increasingly precise control of material properties in both space and time \cite{Salary18, Wang23,Zhou20,Ptitcyn23}, renewing interest in temporal modulation as an additional degree of freedom for electromagnetic design \cite{Engheta21, Engheta23}. Theoretical developments have also broadened the study of time-varying electromagnetic media to include temporally dispersive and nonlocal responses \cite{Solis21,Solis21b,Koutserimpas24}, as well as anisotropic and bianisotropic material systems \cite{VazquezLozano23,Mirmoosa24}. These developments have revealed a broad range of phenomena and functionalities, including temporal reflection and refraction \cite{Vezzoli18,Moussa23,Zhou20,Lustig23}, frequency conversion \cite{Wu20,Wu20b,Moussa23,Lee18}, wave amplification \cite{Koutserimpas18,Wang18}, nonreciprocal propagation \cite{Qin14,Sounas14,Shi17,Chamanara17,Dinc17,Fleury18,Taravati20}, synthetic gauge fields for photons \cite{Fang12,Fang14}, extreme electromagnetic-energy accumulation and trapping \cite{Mirmoosa19,Hecht23}, and temporal analog computing \cite{Rizza22}.

More fundamentally, temporal modulation can relax or circumvent bounds derived under assumptions of time invariance \cite{Shlivinski18,Li19,Yang22,Fritts25,OlguinLopez26,Firestein23,Hayran24}, thereby enlarging the range of electromagnetic responses accessible beyond conventional linear time-invariant systems.

A particularly intriguing manifestation of temporal modulation is the conversion of an initially static electromagnetic state into propagating radiation. This class of phenomena can be traced back to the concept of transition scattering introduced by Ginzburg, in which a stationary electron emits radiation as a consequence of the temporal variation of the surrounding medium \cite{Ginzburg82}. Related mechanisms were subsequently explored in plasma systems \cite{Wilks89,Mori95,Lai96,Esarey96,Murphy06}, including in connection with the generation of high-power electromagnetic radiation. 

More recently, in contrast to earlier studies based on continuously or periodically modulated media, we investigated a transient realization of static-to-dynamic conversion in a dielectric system, showing that an abrupt change in the permittivity of a slab immersed in an electrostatic field can convert initially stored electrostatic energy into propagating radiation \cite{Mencagli2022}.


The analysis in \cite{Mencagli2022}, however, assumed an ideal nondispersive dielectric response and therefore did not account for the temporal memory associated with material polarization. In a dispersive medium, the polarization evolves dynamically and cannot instantaneously adjust to an abrupt change of the material parameters. Consequently, material dispersion affects not only the propagation of the generated electromagnetic transient, but also the way in which the transient is initiated, its early-time evolution, and its modal structure.

In this work, we investigate static-to-dynamic field conversion in a Lorentz-dispersive medium whose oscillator strength is abruptly changed while the system is initially in electrostatic equilibrium. Immediately after the temporal transition, the material polarization retains its pre-switch value and therefore generally differs from the equilibrium polarization associated with the post-switch medium. This polarization mismatch provides the initial excitation for the subsequent transient dynamics. By formulating the problem in the Laplace domain, we derive an analytical representation of the generated field and characterize how material dispersion modifies the slab-mode spectrum.

We show that the dispersive problem possesses a richer spectral structure than its nondispersive counterpart, comprising both remaining slab-mode poles and families of accumulating poles whose accumulation points are associated with the Lorentz polarization dynamics. Material memory also governs the earliest stage of the transient, enforcing a smooth onset of the generated field and setting the velocity of the earliest propagating disturbance through the high-frequency permittivity. In addition, the temporal transition generates broadband spectral content that spans positive-, near-zero-, and negative-permittivity regions of the post-switch material response, giving rise to qualitatively different spatial field distributions.

The remainder of the paper is organized as follows. Section~2 describes the physical configuration and the Lorentz polarization model. Section~3 develops the Laplace-domain formulation of the static-to-dynamic conversion problem. Section~4 derives the corresponding time-domain field solution through Bromwich inversion and a residue representation. Section~5 presents the physical results and interpretation, including the transient dynamics, pole contributions, and spectral and spatial characteristics of the generated field. Finally, the main conclusions are summarized in Sec.~6.

\section{Physical configuration and Lorentz polarization model}
\subsection{Geometry}
\begin{figure}[t!]
    \centering
    \includegraphics[width=\columnwidth]{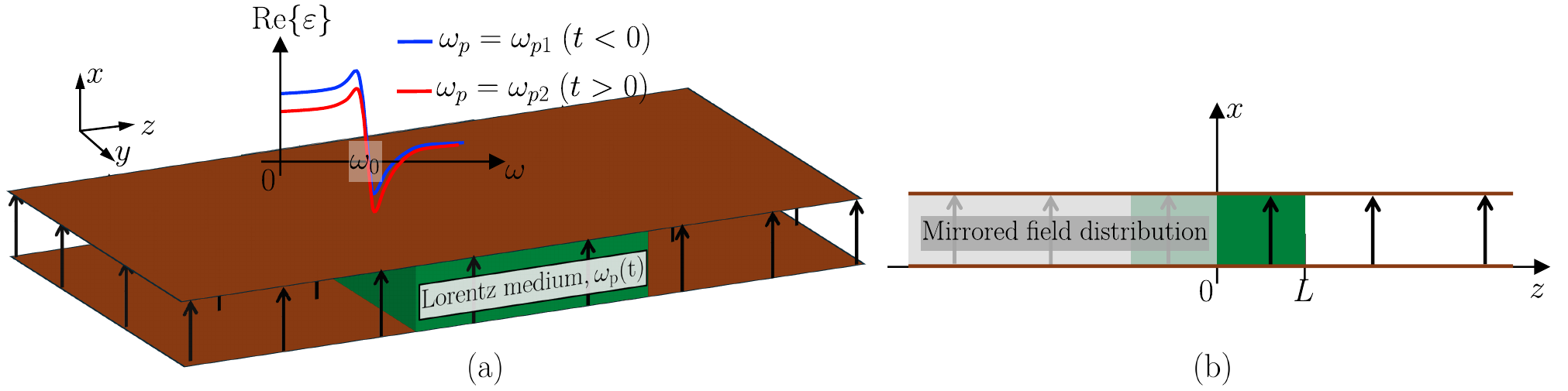}
    \caption{\label{geometry_material}Geometry and material model of the temporally switched dispersive waveguide. (a) A Lorentz-dispersive dielectric slab is embedded between two perfectly conducting parallel plates and initially polarized by a uniform static electric field directed along $x$. At $t=0$, the plasma frequency changes abruptly from $\omega_{p1}$ to $\omega_{p2}$, while the resonance frequency $\omega_{0}$ remains unchanged, thereby modifying the frequency-dependent  permittivity $\varepsilon$. The inset schematically compares the real part of the permittivity before and after the switch. (b) Reduced half-structure obtained by exploiting symmetry about $z=0$; the field distribution in the omitted half is recovered by mirror symmetry.}
\end{figure}

We consider the waveguide geometry shown in Fig.~\ref{geometry_material}. A Lorentz-dispersive dielectric block of length $2L$, occupying $-L<z<L$, is placed at the center of a parallel-plate waveguide with plate separation $d$. The plates are modeled as perfect electric conductors (PECs) and are connected to a DC voltage source that remains connected throughout the entire process. Before the temporal change of the material, the system is in a steady state established by the DC voltage source, which produces an $x$-directed uniform static electric field $\mathbf{E}^\s = E^\s \hat{\mathbf{x}}$ throughout the waveguide. The regions $|z|>L$ are filled with air. The structure is invariant along $y$ and infinite along $z$. Consequently, the electromagnetic response is described by TEM fields with electric and magnetic components directed along $x$ and $y$, respctively. As in \cite{Mencagli2022}, the symmetry about the plane $z=0$ allows us to analyze the reduced half-structure shown in Fig.~\ref{geometry_material}(b). In the reduced problem, the symmetry plane is replaced by a perfect magnetic conductor (PMC) boundary at $z=0$, the Lorentz medium occupies $0<z<L$, and the air region occupies $z>L$.
\subsection{Temporally switched prepolarized Lorentz medium}
The polarization response of the dielectric block is described by the standard Lorentz oscillator model,
\begin{equation}\label{lorentz_pol}
    \ddot{p}_{\LL}(z,t)
    +\gamma \dot{p}_{\LL}(z,t)
    +\omega_0^2 p_{\LL}(z,t)
    =
    \varepsilon_0 \omega_p^2(t) \mathrm{e}_{\LL}(z,t),
\end{equation}
where $p_{\LL}(z,t)$ and $\mathrm{e}_{\LL}(z,t)$ denote the $x$-directed polarization and electric-field components, respectively, $\gamma$ is the damping rate, $\omega_0$ is the resonance frequency, $\varepsilon_0$ is the free-space permittivity, and $\omega_p(t)$ is the time-dependent plasma frequency. The overdot denotes differentiation with respect to time, such that $\dot{p}_\LL=\frac{\partial p_\LL}{\partial t}$ and $\ddot{p}_\LL=\frac{\partial^2 p_\LL}{\partial t^2}$.

Assume that the plasma frequency undergoes an abrupt temporal change at $t=0$, such that $\omega_p(t)=\omega_{p1}$ for $t<0$ and $\omega_p(t)=\omega_{p2}$ for $t>0$. As noted above, the change in $\omega_p$ occurs after the dielectric block has been fully polarized by the static electric field $E_s$. Therefore, before the switching event ($t<0$), the system is in a static steady state. Setting $\dot{p}_\LL=\ddot{p}_\LL=0$ in Eq.~(\ref{lorentz_pol}) gives $P_{1}^\s= \varepsilon_0 \frac{\omega_{p1}^2}{\omega_0^2}E^\s$.

For $t>0$, we decompose the total polarization and electric field into their post-switch static values and dynamic components as
\begin{align}
    p_\LL(z,t)
    &=
    P_{2}^\s+p^{\mathrm{d}}_{\LL}(z,t),\label{decomp_pol}\\
    e_\LL(z,t)
    &=
    E^\s+\mathrm{e}^{\mathrm{d}}_{\LL}(z,t),
    \label{decomp_E_field}
\end{align}
where $P_{2}^\s = \varepsilon_0 \frac{\omega_{p2}^2}{\omega_0^2} E^\s$ is the equilibrium polarization of the post-switch medium under the static electric field $E^\s$. $p_{\LL}^\dd(z,t)$ and $e_{\LL}^\dd(z,t)$ denote the dynamic polarization and electric field, respectively. Eq.~(\ref{lorentz_pol}) combined with Eqs.~(\ref{decomp_pol}) and (\ref{decomp_E_field}), the Lorentz polarization equation reduces to
\begin{equation}\label{dynamic_pol}
    \ddot{p}_{\LL}^\dd
    +\gamma\dot{p}_{\LL}^\dd
    +\omega_0^2p_{\LL}^\dd
    =
    \varepsilon_0\omega_{p2}^2e_{\LL}^\dd.
\end{equation}
Because the total polarization in Eq.~(\ref{decomp_pol}) and its first time derivative must remain continuous across the temporal interface at $t=0$, the initial conditions for the Lorentz polarization equation in Eq.~(\ref{dynamic_pol}) are
\begin{align}
    p_{\LL}^\dd(z,0^+)
    &=
    P_{1}^\s-P_{2}^\s = \varepsilon_0 \frac{\omega_{p1}^2-\omega_{p2}^2}{\omega_0^2} E^\s,\label{dynamic_pol_initial_cond1}\\
    \dot{p}_{\LL}^\dd(z,0^+)
    &=
    0.
    \label{dynamic_pol_initial_cond2}
\end{align}
For compactness, we define the polarization mismatch as $\Delta P\equiv P_{1}^\s-P_{2}^\s$

\section{Laplace-domain formulation}
\subsection{Lorentz polarization relation}
Applying the unilateral Laplace transform \cite{Spiegel1986} to Eq.~(\ref{dynamic_pol}), and using the initial conditions in Eqs.~(\ref{dynamic_pol_initial_cond1}) and (\ref{dynamic_pol_initial_cond2}), we obtain \cite{supp}
\begin{equation}
    P_{\LL}^\dd(z,s)
    =\varepsilon_0
    \frac{\omega_{p2}^2}{A_0(s)}
    E_{\LL}^\dd(z, s)
    +\frac{s+\gamma}{A_0(s)}\Delta P.
\end{equation}
where $A_0(s) = s^2+\gamma s+\omega_0^2$. 
The first term describes the polarization induced by the dynamic electric field, whereas the second arises from the initial conditions of the Lorentz polarization equation and is proportional to the initial polarization displacement $\Delta P=P_{s1}-P_{s2}$. The latter therefore represents the nonequilibrium polarization established by the temporal switch, which drives the subsequent dynamic electromagnetic response.
\subsection{Wave equations and field solutions}
Combining the Laplace-transformed Maxwell equations with the Lorentz polarization relation above leads to the wave equations governing the dynamic electric field. Their detailed derivation is provided in \cite{supp}; here, we state the resulting equations needed to determine the dynamic electric field. Inside the Lorentz medium, the Laplace-domain dynamic electric field satisfies
\begin{equation}
    \frac{\partial^2E_{\LL}^\dd(z, s)}
         {\partial z^2}
    -k_L(s)E_{\LL}^\dd(z, s)
    =-\mu_0\frac{s\omega_0^2}{A_0(s)}\Delta P,
    \label{wave_equation1}
\end{equation}
where
\begin{equation}
    k_{\LL}(s) = k_\mathrm{a}\sqrt{\varepsilon_{2}(s)},
    \label{wave_number}
\end{equation}
\begin{equation}
    \varepsilon_{2}(s) = \varepsilon_\infty \frac{A_1(s)}{A_0(s)},
    \label{eps_post_switch}
\end{equation}
$k_\mathrm{a} = \frac{s}{c_\mathrm{a}}$, $c_\mathrm{a} = \frac{1}{\sqrt{\mu_0\varepsilon_0}}$, $\mu_0$ is the free-space permeability, and $A_1(s) = s^2+\gamma s+\omega_1^2$ with $\omega_1^2 = \omega_0^2+\frac{\omega_{p2}^2}{\varepsilon_\infty}$.
Recalling the definition of $\Delta P$ in Eq.~(\ref{dynamic_pol_initial_cond1}), the inhomogeneous term in Eq.~(\ref{wave_equation1}) originates from the nonzero difference between the polarization immediately after the temporal switch and its new static equilibrium value.

In the air region, where no material polarization is present, the dynamic electric field $E_{\mathrm{a}}^{\mathrm d}$ satisfies the homogeneous equation \cite{supp} with the wavenumber $k_\mathrm{a}$.

Applying the spatial boundary conditions, which include the continuity of the electric and magnetic fields at $z=L$, field symmetry about $z=0$, and the radiation-at-infinity condition, to the general solutions of the wave equations leads to the following electric-field expressions \cite{supp}: 
\begin{equation}\label{E_field_lorentz_L}
    E_{\LL}^\dd(z, s) = E_{\mathrm{part}} \left[ 1 - \frac{k_{\mathrm a} \cosh(k_{\LL} z)}{k_{\mathrm a} \cosh(k_{\LL} L) + k_{\LL} \sinh(k_{\LL} L)} \right],
    \qquad 0\leq z\leq L,
\end{equation}
and
\begin{equation}\label{E_field_air_L}
    E_{\mathrm a}^{\dd}(z,s) = E_{\mathrm{part}} \frac{k_{\LL} \sinh(k_{\LL} L)}{k_\mathrm{a} \cosh(k_{\LL} L) + k_{\LL} \sinh(k_{\LL} L)} e^{-k_\mathrm{a}(z-L)},
    \qquad z>L,
\end{equation}
where
\begin{equation}\label{part_sol}
  E_{\mathrm{part}} = E^\s \frac{\omega_{p1}^2 - \omega_{p2}^2}{s \varepsilon_\infty A_1(s)}.
\end{equation}
The latter is a particular solution arising from the inhomogeneous form of the wave equation in the Lorentz medium \cite{supp}. As expected, the dynamic fields are generated by the temporal switching of the plasma frequency. Indeed, when $\omega_{p2} = \omega_{p1}$, no temporal switch occurs, and $E_{\mathrm{part}}$, $E_{\LL}^{\mathrm d}(z, s)$, and $E_{\mathrm a}^{\mathrm d}(z, s)$ all vanish identically.

Eq.~(\ref{part_sol}) can also be conveniently recast in terms of the pre- and post-switch static permittivities
\begin{equation}\label{part_sol_static}
 E_{\mathrm{part}} = E^\s \left( \frac{\varepsilon_{\s 1}}{\varepsilon_{\s 2}}-1\right)\frac{\omega_1^2}{s A_1(s)}.
\end{equation}
This form is particularly useful for isolating the role of material dispersion, since $\omega_0$ can then be varied while keeping $\varepsilon_{\s 1}$ and $\varepsilon_{\s 2}$ fixed, and it also facilitates comparison with the corresponding dispersionless limit \cite{Mencagli2022}.

\section{Time-domain field solution}
To obtain the electric fields in the time domain [Eqs.~(\ref{E_field_lorentz_L}) and (\ref{E_field_air_L})], we evaluate the inverse
Laplace transform through the Bromwich integral \cite{Spiegel1986},
\begin{equation}
    \mathrm{e}_{\LL}^\dd(z, t)
    =
    \frac{1}{2\pi i}
    \int_{\sigma_\B-i\infty}^{\sigma_\B+i\infty}
    E_{\LL}^\dd(z, s)e^{st}\,ds,\qquad t>0,
\end{equation}
where the vertical integration line $\operatorname{Re}(s)=\sigma_\B$ lies to the right of all singularities of the integrand. In the remainder of the analysis, we focus on the
electric field inside the Lorentz medium, since the field in the air region exhibits an analogous analytic structure and can be treated using the same procedure.

Evaluating the Bromwich integral by contour deformation and the residue theorem requires a careful characterization of the analytic structure of the Laplace-domain field. Beyond providing the mathematical basis for the inversion, this analysis offers a physical interpretation of the transient response by identifying the slab-mode poles and the dispersion-induced slab-mode poles that contribute to the field. We therefore examine the analyticity and pole structure of the Laplace-domain field in the following section before deriving its residue representation.
\subsection{Analyticity of the Bromwich integrand and interpretation of the pole structure}
The electric field inside the Lorentz medium is recovered from the Bromwich integral
\begin{equation}
\mathrm{e}_{\LL}^\dd(z,t) = \frac{E^\s}{2\pi i} \left( \frac{\varepsilon_{\s 1}}{\varepsilon_{\s 2}}-1\right)
    \int_{\sigma_\B-i\infty}^{\sigma_\B+i\infty}
    \Phi(s,z,t)\,ds,\qquad t>0,
\end{equation}
where the Bromwich integrand is defined as
\begin{equation}
    \Phi(s,z,t)
    =
    \frac{\omega_1^2}{sA_1(s)}
    \frac{N(s,z)}{D(s)}
    e^{st},
    \label{bromwich_int}
\end{equation}
with
\begin{equation}
D(s) = \cosh\!\left[k_{\LL}(s)L\right] +\frac{k_{\LL}(s)}{k_\mathrm{a}(s)}\sinh\!\left[k_{\LL}(s)L\right]
\label{D_def}
\end{equation}
and
\begin{align}
N(s,z) = D(s)-\cosh\!\left[k_{\LL}(s)z\right].
\label{N_def}
\end{align}
Hence, the analytic structure governing the Bromwich inversion is determined by $\Phi(s,z,t)$. For any finite $t$, the exponential factor $e^{st}$ is entire and nonvanishing throughout the finite $s$ plane. It therefore introduces no additional finite singularities and leaves unchanged the locations and orders of those inherited from the Laplace-domain field. Including $e^{st}$ in the definition of $\Phi(s,z,t)$ is nevertheless convenient, because the residue at each pole directly leads to the complete time-dependent modal contribution. Moreover, the exponential factor controls the large-$|s|$ behavior of the integrand and is therefore essential to the contour-deformation analysis.

To analyze the analyticity of $\Phi(s,z,t)$, we first examine its dependence on the square-root branch of $k_{\LL}(s)$ [Eq.~(\ref{wave_number})]. It can be readily shown that the complete Bromwich integrand is invariant under the transformation $k_{\LL}(s)\mapsto -k_{\LL}(s)$. Therefore, the branch discontinuity associated with $k_{\LL}(s)$ does not produce a branch-cut contribution to the inverse Laplace transform.

We next consider the apparent singularities at $s=0$ and at the roots of $A_1(s)$. It can be shown that the corresponding zeros of the numerator cancel the explicit factors $1/s$ and $1/A_1(s)$, respectively. Hence, both $s=0$ and the roots of $A_1(s)$ are removable singularities of $\Phi(s,z,t)$ and do not generate independent contributions to the inverse Laplace transform. 

At the roots of $A_0(s)$, the wavenumber $k_{\LL}(s)$ [Eq.~(\ref{wave_number})] becomes singular. These roots are not isolated singularities of the complete Bromwich integrand. Instead, infinitely many isolated zeros of $D(s)$ accumulate toward each of them \cite{supp}. The corresponding poles of the complete spatial field will hereafter be referred to as accumulating poles. Each accumulating pole is isolated and therefore possesses a well-defined residue. In contrast, the roots of $A_0(s)$ are nonisolated singularities, since they constitute accumulation points of infinitely many zeros of $D(s)$. Consequently, no residue can be assigned directly to these roots, and they contribute no additional term beyond the residues of the isolated poles approaching them. A detailed derivation of the local pole asymptotics and of the vanishing contribution from admissible pole-avoiding contours shrinking around these roots is provided in Ref.~\cite{supp}.

In addition to the accumulating-pole families described above, $D(s)$ possesses other isolated zeros that remain separated from the roots of $A_0(s)$. The corresponding poles of the complete field will be referred to as remaining poles. Like the accumulating poles, they represent modes of the finite Lorentz slab coupled to the surrounding air region.

The complete pole spectrum therefore consists of two families of slab modes distinguished by their relation to the Lorentz singularities. The accumulating poles form infinite sequences that approach the roots of \(A_0(s)\), whereas the remaining poles stay separated from these accumulation points. The material dispersion is responsible for the emergence of the accumulation-point structure, while the slab geometry, interface conditions, and material response together determine the locations and residues of all isolated poles. The transient field is consequently governed by the collective contribution of both the accumulating and remaining poles.

The analytic structure relevant to the Bromwich inversion can therefore be summarized as follows: \(s=0\) and the roots of \(A_1(s)\) are removable singularities; the roots of \(A_0(s)\) are nonisolated accumulation points of the accumulating-pole families; and the isolated zeros of \(D(s)\) corresponding to genuine poles provide the residue contributions to the time-domain field. These results form the basis for the contour deformation and residue representation developed in the following subsection.

\subsection{Bromwich inversion and residue representation}
Having characterized the analytic structure of the Bromwich integrand, we now evaluate the inverse Laplace transform by deforming the Bromwich contour into the left half of the complex $s$ plane. For $t>0$, as shown in Ref.~\cite{supp}, the contribution from the large-$|s|$ closing contour vanishes. The contributions from admissible pole-avoiding contours shrinking around the roots of $A_0(s)$ also vanish. The time-domain electric field is therefore determined entirely by the residues of the isolated poles of $\Phi(s,z,t)$. Because the material and geometrical parameters are real, the nonreal poles occur in complex-conjugate pairs, while a single isolated real pole $s_0$ is present. Denoting by $\mathcal P_+$ the poles in the upper half of the complex $s$ plane, the dynamic field can therefore be written as
\begin{equation}
{\mathrm e}_{\LL}^{\mathrm d}(z,t)
=
E^\s
\left(
\frac{\varepsilon_{\s 1}}{\varepsilon_{\s 2}}-1
\right)
\left[
r_0(z)e^{s_0t}
+
2\operatorname{Re}
\left\{
\sum_{s_{n}\in\mathcal{P}_+}
r_{n}(z)e^{s_nt}
\right\}
\right],
\qquad t>0,
\label{modal_expansion}
\end{equation}
where
\begin{equation}
r_n(z)=
\frac{\omega_1^2 N(s_n,z)}
{s_n A_1(s_n)D'(s_n)}.
\label{modal_coefficient}
\end{equation}
The pole set $\mathcal P_+$ contains both the accumulating and remaining poles introduced above.
%
%
%

%
%

The isolated poles entering Eq.~(\ref{modal_expansion}) correspond to modes of the complete finite Lorentz slab, with their locations determined jointly by the material dispersion, slab thickness, and interface conditions. They are separated into two groups: accumulating poles, which form families approaching the roots of $A_0(s)$, and remaining poles, which stay separated from these accumulation points. Both groups consist of genuine isolated poles and therefore contribute through their residues. In contrast, the roots of $A_0(s)$ are nonisolated accumulation points rather than poles of the complete field and introduce no additional terms in the residue representation.

%
%

The prefactor in Eq.~(\ref{modal_expansion}) sets the overall excitation strength associated with the change in the equilibrium polarization induced by the temporal switch, whereas the coefficients $r_n(z)$ determine the spatially dependent contribution of each isolated slab mode to the transient field.

\section{Physical results and interpretation}
Having established the time-domain solution in terms of the poles and residues of the Bromwich integrand, we now examine the physical behavior of the generated transient. We first validate the residue representation and assess its stability and numerical convergence. We then investigate the early-time response and the role of material memory, examine the recovery of the dispersionless limit and the response in different dispersive regimes, finally identify the contributions of the different pole families.
\subsection{Validation of residue representation}
For the realistic material parameters considered here, all isolated poles contributing to the residue representation satisfy $\operatorname{Re}\{s_n\}<0$, indicating that the post-switch modal dynamics are temporally stable. The detailed pole distribution and its physical interpretation are discussed in Sec. 5.3.

\begin{figure}[t!]
    \centering
    \includegraphics[width=\columnwidth]{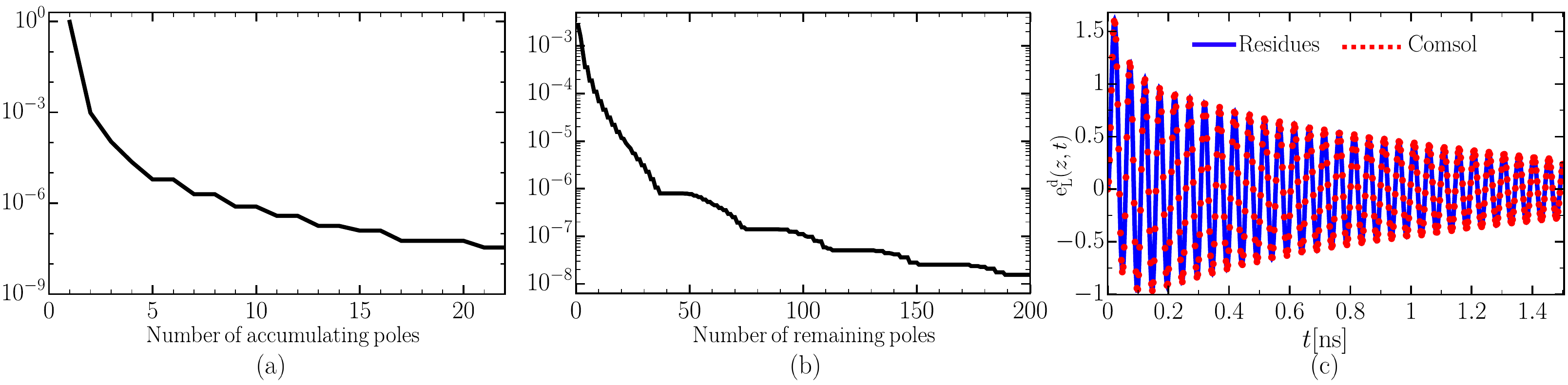}
    \caption{\label{validation}Validation of the residue representation. Relative truncation-error envelope of the residue expansion for (a) the accumulating poles and (b) the remaining poles, as a function of the number of included conjugate pole pairs. The systematic decrease of the error envelopes demonstrates numerical convergence of both contributions to the residue representation. (c) Time-domain dynamic electric field reconstructed from the converged residue expansion and obtained independently from a COMSOL Multiphysics\textsuperscript{\textregistered} transient simulation, showing excellent agreement. All results are evaluated at $z_0=L/2$, while the convergence results in panels (a) and (b) correspond to $t=0.5\,\mathrm{ps}$. The parameters are $\omega_0/(2\pi)=20\,\mathrm{GHz}$, $L=3\, \mathrm{mm}$, $\gamma=\omega_0/100$, $\varepsilon_{\infty}=3.9$, $E^\s=1$, $\varepsilon_{\s 1}=8$, and $\varepsilon_{\s 2}=4$}
\end{figure}

Temporal stability, however, does not by itself establish convergence of the residue expansion, which also depends on the behavior of the modal contributions as increasingly higher-order poles are included. Figure~\ref{validation} examines the numerical convergence of the residue representation. The single real pole is included separately, while the nonreal poles occur in complex-conjugate pairs; accordingly, each increment in Figs.~\ref{validation}(a) and \ref{validation}(b) corresponds to the inclusion of one additional conjugate pole pair. The relative truncation-error envelope is evaluated separately for the accumulating and remaining poles, and its systematic decrease demonstrates convergence of both contributions. The residue-based solution is further validated against an independent time-domain simulation performed using COMSOL Multiphysics\textsuperscript{\textregistered} [Fig.~\ref{validation}(c)].
\subsection{Early-time dynamics and material memory}
\begin{figure}[t!]
    \centering
    \includegraphics[scale=0.3]{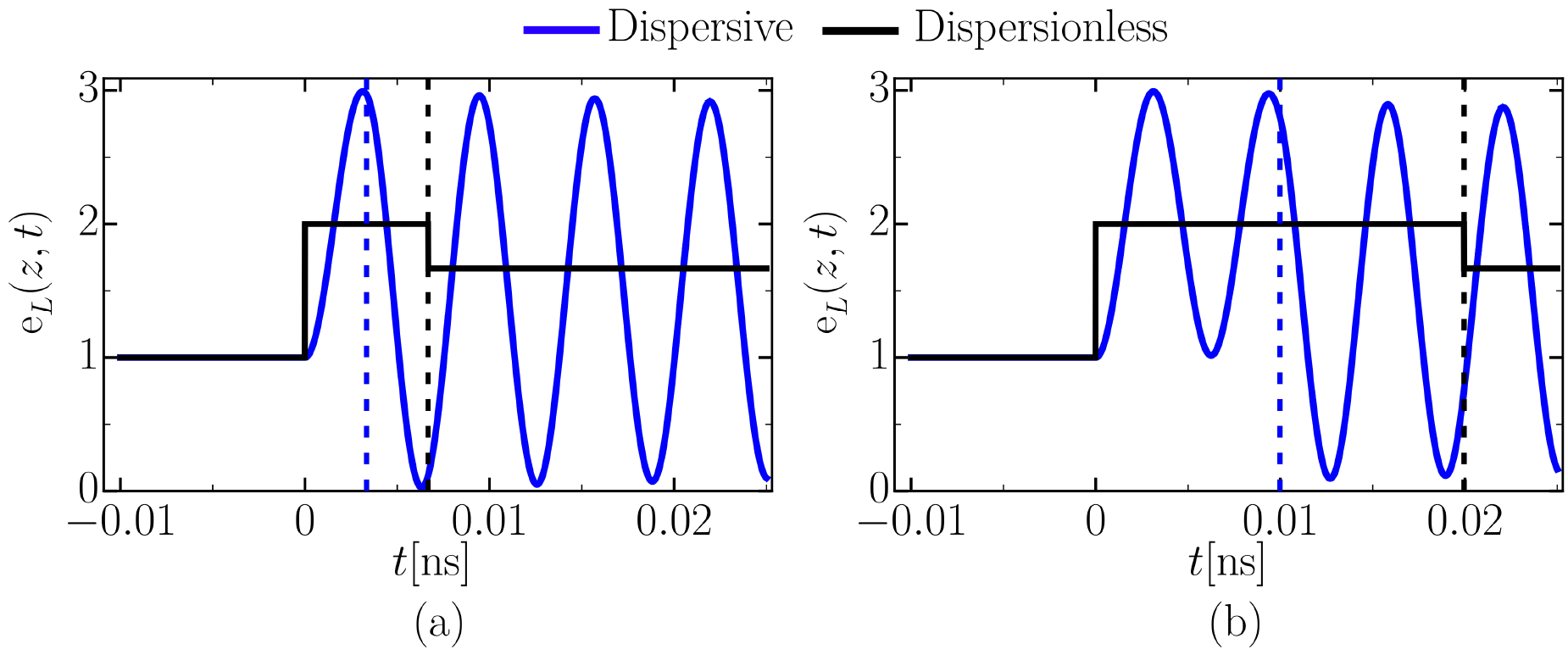}
    \caption{\label{early_time_memory} Comparison between the dispersive solution (blue) and the dispersionless early-time approximation (black) for the total electric field $e_\LL(z,t)$ at two observation positions: (a) $z=2\, \mathrm{mm}$ and (b) $z=0.01\, \mathrm{mm}$. The vertical dashed lines indicate the onset of the interface-dependent contributions at each observation point. The parameters are $\omega_0/(2\pi)=80\,\mathrm{GHz}$, $L=3\, \mathrm{mm}$, $\gamma=\omega_0/100$, $\varepsilon_{\infty}=1$, $E^\s=1$, $\varepsilon_{s1}=8$, and $\varepsilon_{s2}=4$.}
\end{figure}

To investigate how material dispersion affects the initial stage of the generated transient, we consider the early-time behavior of the dynamic electric field inside the Lorentz slab. This regime can be analyzed from the large-$|s|$ asymptotic behavior of the Laplace-domain solution, which provides a compact approximation before higher-order reflections significantly contribute to the field. Retaining the local response generated by the temporal switch and the first contribution associated with the slab interface, the dynamic electric field can be approximated as
\begin{equation}\label{early_time_field}
{\mathrm e}_{\LL}^{\mathrm d}(z,t)\simeq
E^\s\left(\frac{\varepsilon_{\s 1}}{\varepsilon_{\s 2}}-1\right)
\left\{ [1-\phi(t)]u(t)
-\frac{1}{1+\sqrt{\varepsilon_\infty}}
[1-\phi(t-\tau)]u(t-\tau)\right\}
\end{equation}
where
\begin{equation}
\phi(t)=
e^{-\gamma t/2}
\left[
\cos\left(\omega_1 t\right)
+\frac{\gamma}{2\omega_1}
\sin\left(\omega_1 t\right)
\right],
\end{equation}
$\tau=\frac{L-z}{c_\infty}$, $c_\infty=\frac{c_a}{\sqrt{\varepsilon_\infty}}$, and $u(\cdot)$ denotes the Heaviside unit-step function. Inspecting Eq.~(\ref{early_time_field}), two distinct signatures of dispersion can be identified, both of which are illustrated in Fig.~\ref{early_time_memory}. Before the arrival of any interface-dependent contribution, $0<t<\tau$, the field evolution is governed solely by the local material response. In particular, the dynamic field exhibits the short-time behavior ${\mathrm e}_L^{\mathrm d}(z,t)\propto t^2$. Accordingly, as shown by the blue curves in Fig.~\ref{early_time_memory}, the total field evolves continuously from its pre-switch value, in marked contrast with the instantaneous jump predicted by the dispersionless model (black curves). This initial smooth evolution directly reflects the finite memory of the Lorentz polarization, which cannot adjust instantaneously to the post-switch equilibrium.
A second signature of dispersion is associated with the delayed, interface-dependent contribution. Its onset occurs at $\tau=\frac{L-z}{c_\infty}$, $c_\infty=\frac{c_{\mathrm a}}{\sqrt{\varepsilon_\infty}}$. The corresponding arrival times are indicated by the vertical dashed lines in Fig.~\ref{early_time_memory}. Comparing the two observation positions shows the expected \(z\)-dependent shift of these arrival times, confirming that the earliest interface-mediated disturbance propagates through the Lorentz medium at the front velocity \(c_\infty\), determined by the high-frequency permittivity \(\varepsilon_\infty\), rather than by the static permittivity. Beyond these characteristic times, the total field begins to contain contributions mediated by the interfaces, and the purely local early-time description no longer applies.

\subsection{Modal composition of the transient}
\begin{figure}[t!]
    \centering
    \includegraphics[scale=0.3]{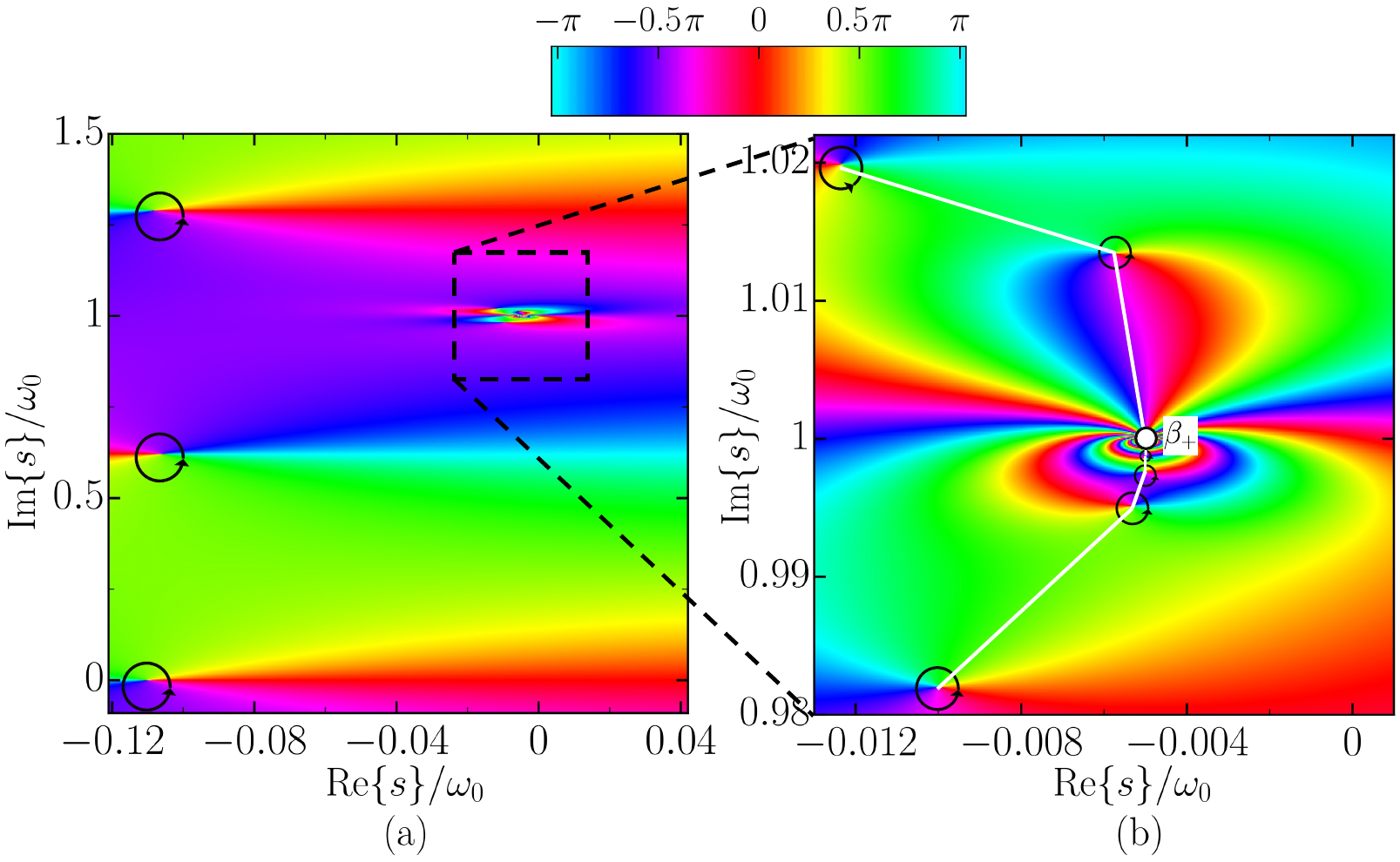}
    \caption{\label{pole_structure} Pole structure of the Bromwich integrand [Eq.~(\ref{bromwich_int})]. Phase colormap of the denominator $D(s)$ [Eq.~(\ref{D_def})] in the complex $s$-plane. The phase vortices identify zeros of $D(s)$, which correspond to isolated poles of the Bromwich. (a) Global view of the pole distribution. (b) Magnified view of the region near the upper accumulation point $\beta_+$, showing the sequence of dispersion-induced slab-mode poles approaching the Lorentz singularity. Black circles mark the numerically determined zeros of $D(s)$. The parameters are $\omega_0/(2\pi)=40\,\mathrm{GHz}$, $L=3\, \mathrm{mm}$, $\gamma=\omega_0/100$, $\varepsilon_{\infty}=3.9$, $\varepsilon_{s1}=8$, and $\varepsilon_{s2}=4$.}
\end{figure}

The residue representation provides a direct means of separating the contributions of the different portions of the pole spectrum to the generated transient. Fig.~\ref{pole_structure}(a) shows the phase of the denominator $D(s)$ [Eq.~(\ref{D_def})] in the complex $s$-plane, with its zeros identified by the corresponding phase vortices. These zeros determine the isolated poles of the Bromwich integrand [Eq.~(\ref{bromwich_int})]. In addition to the remaining slab-mode poles, the spectrum contains the dispersion-induced slab-mode families that accumulate toward the Lorentz singularities $\beta_\pm$ \cite{supp}. The magnified view in Fig.~\ref{pole_structure}(b) clearly reveals the progressive clustering of these poles in the vicinity of $\beta_+$; the corresponding sequence around $\beta_-$ follows by complex conjugation. Thus, although both sets of poles belong to the modes of the complete finite Lorentz slab, the accumulation families constitute a distinct spectral signature introduced by material dispersion.

\begin{figure}[t!]
    \centering
    \includegraphics[width=\columnwidth]{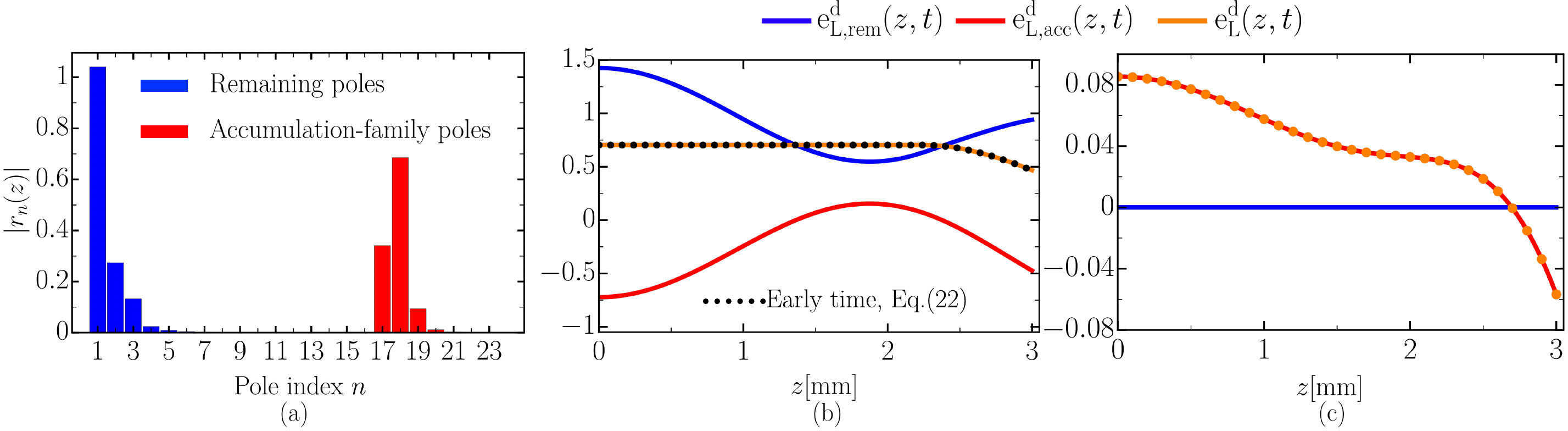}
    \caption{\label{modal_decomposition} Modal contributions to the dynamic electric field. (a) Magnitudes of the residue coefficients associated with the remaining slab-mode poles (blue) and the dispersion-induced slab-mode poles (red), evaluated at $z=0$. Spatial distribution of the dynamic electric field at (b) $t=50\,\mathrm{ps}$ and (c) $t=0.5\,\mathrm{ns}$, showing separately the contributions $\mathrm{e}_{\LL,\mathrm{rem}}^{\mathrm d}$ (blue) and $\mathrm{e}_{\LL,\mathrm{acc}}^{\mathrm d}$ (red), together with their sum $\mathrm{e}_{\LL}^{\mathrm d}$ (orange). In (b), the black dotted curve denotes the early-time approximation. The parameters are $\omega_0/(2\pi)=40\,\mathrm{GHz}$, $L=3\, \mathrm{mm}$, $\gamma=\omega_0/100$, $\varepsilon_{\infty}=3.9$, $\varepsilon_{s1}=8$, and $\varepsilon_{s2}=4$.}
\end{figure}

To assess how the two portions of the spectrum contribute to the transient, we decompose the dynamic electric field as $\mathrm{e}_{\LL}^{\mathrm d}(z,t) = \mathrm{e}_{\LL,\mathrm{rem}}^{\mathrm d}(z,t) + \mathrm{e}_{\LL,\mathrm{acc}}^{\mathrm d}(z,t)$, where $\mathrm{e}_{\LL,\mathrm{rem}}^{\mathrm d}(z,t)$ contains the contribution of the remaining slab-mode poles and $\mathrm{e}_{\LL,\mathrm{acc}}^{\mathrm d}(z,t)$ that of the dispersion-induced pole families accumulating toward $\beta_\pm$. Fig.~\ref{modal_decomposition}(a) shows the magnitudes of the corresponding residue coefficients at $z=0$. For both families, the modal weights decrease rapidly with pole order, although their distributions are different. This illustrates that the pole locations determine the available dynamics, whereas the residues determine the strength with which the corresponding modes are excited.

The spatial field decomposition at $t=50\,\mathrm{ps}$, shown in Fig.~\ref{modal_decomposition}(b), reveals that both pole families contribute appreciably during the early stage of the transient. Their contributions partially compensate each other, and their sum closely follows the early-time approximation of Eq.~(\ref{early_time_field}), shown by the black dotted curve. At the later time $t=0.5\,\mathrm{ns}$, Fig.~\ref{modal_decomposition}(c), the relative balance changes substantially: the contribution associated with the remaining slab-mode poles has strongly decayed, whereas the accumulation-family contribution nearly coincides with the total dynamic field. This behavior is consistent with the pole distribution discussed above. The accumulation-family poles approach $\beta_\pm$, whose real part is $-\gamma/2$, and therefore lie closer to the imaginary axis than the more strongly damped remaining slab-mode poles. Consequently, although the two families both participate in building the early transient, the more slowly decaying accumulation-family modes become dominant at later times for the parameters considered here.

\subsection{Recovery of the dispersionless limit}
For fixed static permittivities, the resonance frequency $\omega_0$ controls the degree of dispersion experienced over the frequency range relevant to the transient. Increasing $\omega_0$ progressively shifts the material resonance away from this spectral region and drives the system toward the corresponding dispersionless limit.

Eq.~(\ref{early_time_field}) accurately describes the earliest high-frequency wavefront for any finite Lorentz resonance frequency, which propagates at $c_\infty$. However, because this approximation is obtained from the large-$|s|$ behavior at fixed $\omega_0$, it does not uniformly recover the dispersionless propagation velocity as $\omega_0\rightarrow\infty$. The dispersionless limit must instead be taken at the level of the complete Laplace-domain solution, for which the post-switch permittivity approaches $\varepsilon_{s2}$ at every finite $s$, yielding the propagation velocity $c_a/\sqrt{\varepsilon_{s2}}$.

The evolution of the modal spectrum provides a complementary interpretation of this limit. As $\omega_0$ increases, the dispersion-induced pole families are displaced toward progressively higher frequencies, together with the corresponding accumulation points. Since the material damping rate $\gamma$ is kept fixed, these modes are not increasingly attenuated as $\omega_0$ grows. Rather, their contribution is shifted toward increasingly rapid oscillations, whose collective contribution to the finite-time field becomes progressively less important through modal cancellation.

\begin{figure}[t!]
    \centering
    \includegraphics[scale=0.25]{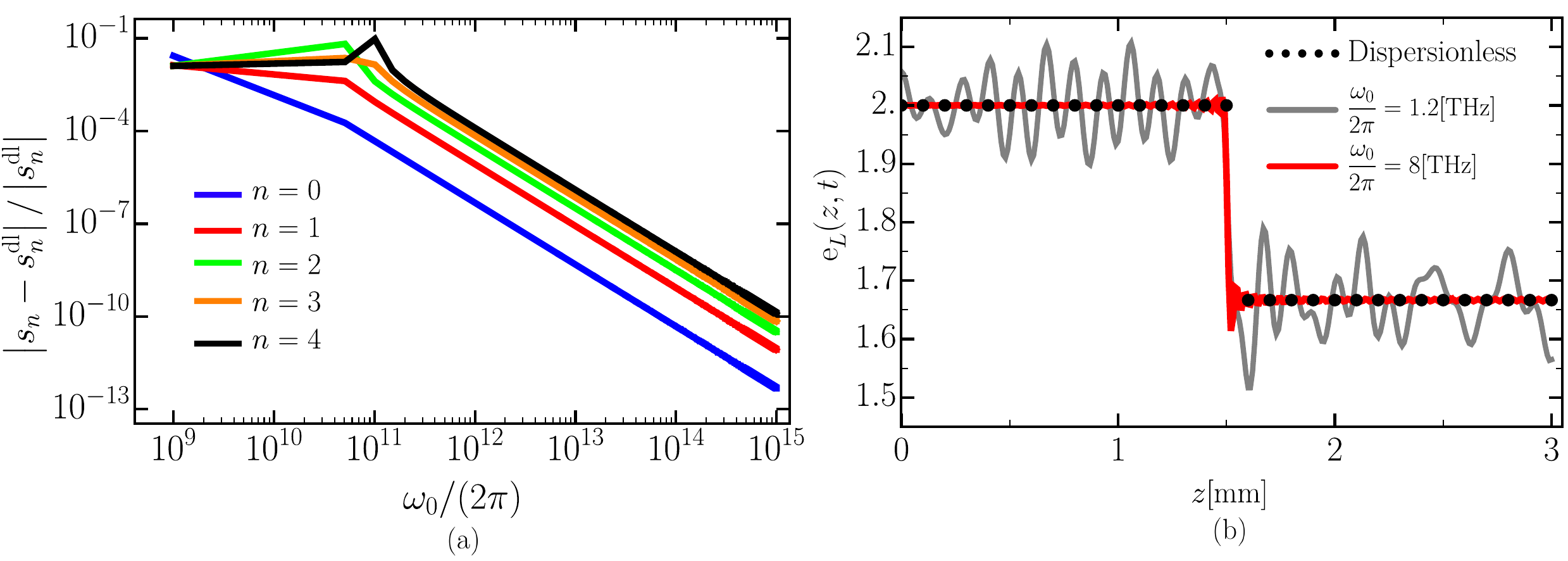}
    \caption{\label{recovery_displess} Recovery of the dispersionless limit. (a) Relative deviation of the first five remaining slab-mode poles $s_n$ from their dispersionless counterparts $s_n^{\mathrm{dl}}$ as a function of the Lorentz resonance frequency \(\omega_0/(2\pi)\). The pole positions progressively approach the dispersionless spectrum as $\omega_0$ increases. (b) Spatial distribution of the dynamic electric field for $\omega_0/(2\pi)=1.2\,\mathrm{THz}$ and $8\,\mathrm{THz}$, compared with the analytical dispersionless solution. The field approaches the dispersionless response as the material resonance is shifted to higher frequencies. The damping rate is kept fixed at $\gamma=\omega_{0,\mathrm{ref}}/100$, with $\omega_{0,\mathrm{ref}}/(2\pi)=40\,\mathrm{GHz}$. The remaining parameters are $L=3\,\mathrm{mm}$, $\varepsilon_\infty=3.9$, $\varepsilon_{s1}=8$, and $\varepsilon_{s2}=4$.}
\end{figure}

At the same time, the remaining slab-mode poles approach the poles of the corresponding dispersionless problem. To quantify this convergence, Fig.~\ref{recovery_displess}(a) shows the relative deviation for the first five remaining slab-mode poles. At sufficiently large $\omega_0$, all five curves exhibit an approximately parallel algebraic decay. This behavior follows directly from the large-$\omega_0$ expansion of the Lorentz permittivity, which implies that, for each fixed remaining mode, $    s_n-s_n^{\mathrm{dl}} =
\mathcal{O}\!\left(\omega_0^{-2}\right)$
The numerical results in Fig.~\ref{recovery_displess}(a) are consistent with this asymptotic scaling.

The same convergence is observed directly in the spatial field distribution. Fig.~\ref{recovery_displess}(b) compares the dynamic electric field at a fixed observation time for two values of $\omega_0$ with the analytical dispersionless  \cite{Mencagli2022}. For $\omega_0/(2\pi)=1.2\,\mathrm{THz}$, residual dispersive oscillations are still clearly visible throughout the slab. When the resonance frequency is increased to $\omega_0/(2\pi)=8\,\mathrm{THz}$, these oscillations are strongly reduced and the field closely follows the dispersionless spatial profile, including the location and magnitude of the propagating discontinuity. Thus, the approach to the dispersionless regime is manifested both spectrally, through the convergence of the remaining slab-mode poles, and in the time-domain field, which approaches the corresponding nondispersive response.
\subsection{Spectral regimes and transient-field evolution}
\begin{figure}[ht!]
    \centering
    \includegraphics[scale=0.25]{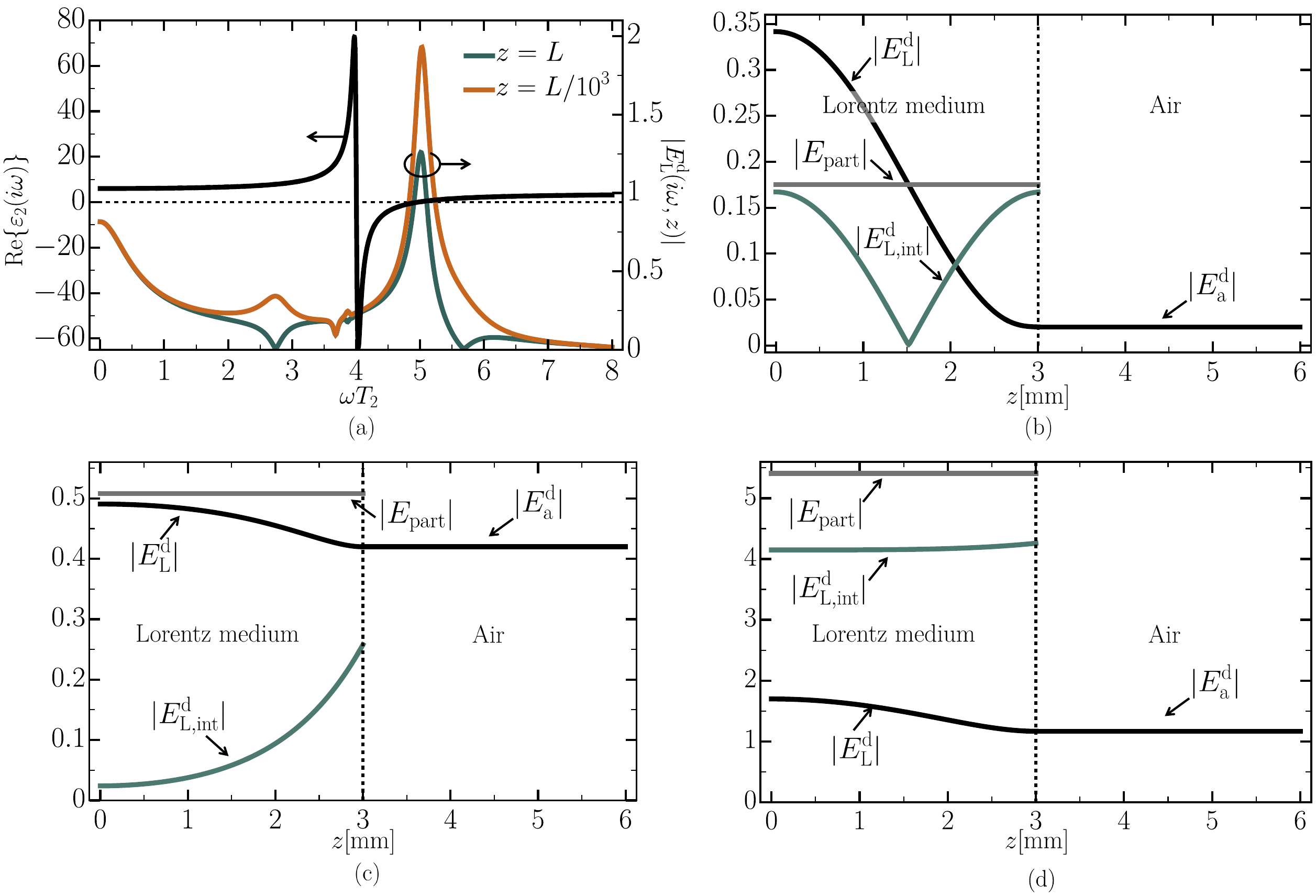}
    \caption{\label{perm_regimes} Spectral and spatial characteristics of the temporally generated field across different permittivity regimes. (a) Real part of the post-switch Lorentz permittivity, \(\operatorname{Re}\{\varepsilon_2(i\omega)\}\) (black, left axis), together with the magnitude of the dynamic-field spectrum evaluated at \(z=L\) and \(z=L/10^3\) (right axis). The marked frequencies correspond to representative positive-, negative-, and near-zero-permittivity regimes. (b) Positive-permittivity regime, showing the particular contribution \(|E_{\mathrm{part}}|\), the interface-dependent contribution \(|E_{\LL,\mathrm{int}}^{\mathrm d}|\), the total field \(|E_{\LL}^{\mathrm d}|\), and the corresponding air field \(|E_{\mathrm a}^{\mathrm d}|\). (c) Negative-permittivity regime, where the interface-dependent contribution exhibits evanescent-like attenuation into the Lorentz medium. (d) Near-zero-permittivity regime, where the particular and interface-dependent contributions are nearly spatially uniform and their coherent superposition yields a substantially smaller total field. The vertical dotted line denotes the Lorentz-medium--air interface at \(z=L\). The parameters are \(\omega_0/(2\pi)=26\,\mathrm{GHz}\), \(L=3\,\mathrm{mm}\), \(\gamma=2\pi40\,[\mathrm{GHz}]/100\), \(\varepsilon_\infty=3.9\), \(\varepsilon_{\s1}=8\), and \(\varepsilon_{\s2}=4\).}
\end{figure}

The temporal switch generates broadband spectral content that spans qualitatively different portions of the post-switch Lorentz dispersion. This behavior is illustrated in Fig.~\ref{perm_regimes}(a), where the real part of the post-switch permittivity is shown together with the magnitude of the generated electric-field spectrum at two representative positions inside the Lorentz region. The spectrum extends across positive-, near-zero-, and negative-permittivity regions, with its detailed frequency dependence shaped by the dispersive material response and the finite-slab configuration. For the parameters considered here, a pronounced spectral maximum occurs in the vicinity of the frequency at which \(\operatorname{Re}\{\varepsilon_2\}\) crosses zero.

To interpret the spatial behavior associated with different portions of the spectrum, we refer to the Laplace-domain field derived above for the Lorentz , [Eq.~(\ref{E_field_lorentz_L})]. In addition to the previously defined spatially uniform particular solution [Eq.~\ref{part_sol_static}], we denote by $E_{\LL,\mathrm{int}}^{\mathrm d}(z,\omega)$ the interface-dependent contribution, which contains the spatial dependence associated with the finite-slab response and the boundary conditions at the Lorentz-medium--air interface. The coherent superposition of these two contributions gives the total field inside the Lorentz medium. As required by the boundary conditions, the total electric field is continuous across the Lorentz-medium--air interface at \(z=L\), irrespective of the permittivity regime considered. In the lossless air region, each spectral component propagates without attenuation; consequently, the magnitude of \(E_{\mathrm a}^{\mathrm d}\) is independent of \(z\), while its phase varies with propagation distance.
Fig.~\ref{perm_regimes}(b) shows a representative frequency in the positive-permittivity regime, where $\operatorname{Re}\{\varepsilon_2\}>0$ and the material loss is weak. When evaluated along the physical-frequency axis, $k_{\LL}$ is predominantly imaginary, giving the interface-dependent contribution $E_{\LL,\mathrm{int}}^{\mathrm d}$ an approximately oscillatory, standing-wave-like spatial profile inside the Lorentz slab. At the selected frequency, its magnitude is comparable to that of the spatially uniform particular solution, so that their coherent superposition produces a strongly position-dependent total field. The total field remains continuous at $z=L$ and couples to a propagating wave of constant magnitude in the lossless air region.

A qualitatively different behavior occurs in the negative-permittivity region, as shown in Fig.~\ref{perm_regimes}(c). At the selected frequency, \(\operatorname{Re}\{\varepsilon_2\}<0\) while \(|\operatorname{Im}\{\varepsilon_2\}|\) remains substantially smaller than \(|\operatorname{Re}\{\varepsilon_2\}|\). The interface-dependent contribution \(E_{\LL,\mathrm{int}}^{\mathrm d}\) consequently exhibits an evanescent-like spatial variation, with its magnitude decreasing away from the Lorentz-medium--air interface and into the Lorentz medium. For the parameters considered here, the attenuation is therefore predominantly associated with the negative real part of the permittivity rather than with material absorption. Once coupled into the lossless air region, however, the same spectral component becomes propagating and its magnitude remains constant with propagation distance.

\begin{figure}[t!]
    \centering
    \includegraphics[scale=0.25]{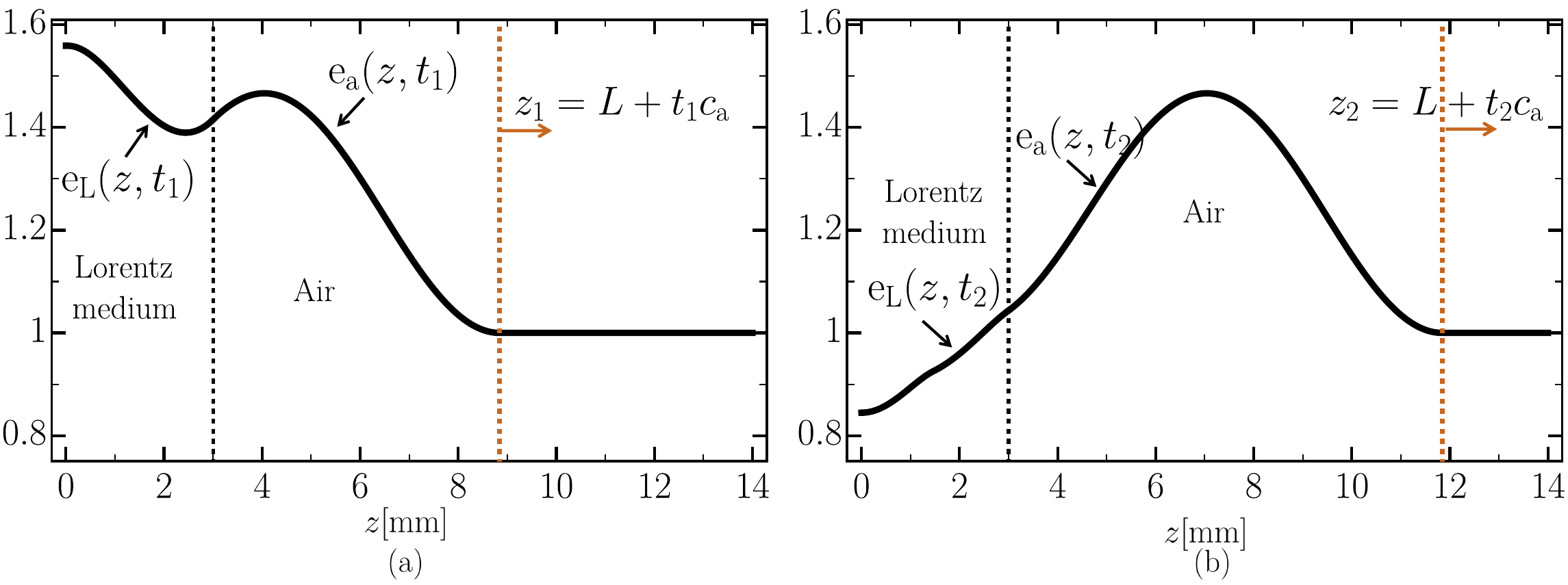}
    \caption{\label{time_evolution} Time-domain evolution of the generated transient. Spatial distribution of the electric field at two representative times, (a) $t=t_1=19.5\,\mathrm{ps}$ and (b) $t=t_2=29.5\,\mathrm{ps}$, showing the field inside the Lorentz medium, ${\mathrm e}_{\LL}(z,t)$, and the outgoing field in air, ${\mathrm e}_{\mathrm a}(z,t)$. The black dotted line marks the Lorentz-medium–air interface at $z=L$, where the electric field remains continuous. The orange dotted line indicates the causal wavefront in air, $z_{1,2}=L+c_{\mathrm a} t_{1,2}$; ahead of the wavefront the field remains at its undisturbed static value. The comparison illustrates the temporal evolution of the internal transient and its propagation away from the interface into the air region. The vertical dotted line denotes the Lorentz-medium–air interface at $z=L$. The parameters are $\omega_0/(2\pi)=26\,\mathrm{GHz}$, $L=3\, \mathrm{mm}$, $\gamma=2 \pi 40[\mathrm{GHz}]/100$, $\varepsilon_{\infty}=3.9$, $\varepsilon_{\s 1}=8$, and $\varepsilon_{\s 2}=4$.}
\end{figure}

The near-zero-permittivity regime is illustrated in Fig.~\ref{perm_regimes}(d), at a frequency close to the zero crossing of $\operatorname{Re}\{\varepsilon_2\}$. In this regime, the reduction of $|k_\LL|$ leads to weak spatial variation of the interface-dependent contribution across the slab. Its magnitude is therefore nearly uniform, as is that of the previously defined particular solution. Although both individual contributions are comparatively large, their coherent superposition produces a substantially smaller total field throughout much of the Lorentz region. The total field remains finite and continuous with the field in air at $z=L$.

The frequency-domain analysis of Fig.~\ref{perm_regimes} describes the spatial character of the individual spectral components generated by the temporal switch. Their coherent superposition determines the time-domain evolution shown in Fig.~\ref{time_evolution}. The spatial field distribution is plotted at two representative times, \(t_1=19.5\,\mathrm{ps}\) and \(t_2=29.5\,\mathrm{ps}\). At both times, the field is continuous across the Lorentz-medium--air interface at \(z=L\), while the field inside the Lorentz medium evolves as the material polarization relaxes from its initial nonequilibrium state.

In the air region, the generated disturbance propagates away from the interface at the velocity \(c_a\). Accordingly, at time \(t_i\) the leading edge of the transient is located at \(z_i=L+c_at_i\), as indicated by the orange dotted lines in Fig.~\ref{time_evolution}. Ahead of this causal wavefront, the field remains at its undisturbed static value. Comparing Figs.~\ref{time_evolution}(a) and \ref{time_evolution}(b) shows the advance of the radiated transient into the air region together with the simultaneous evolution of the field inside the Lorentz medium. The figure therefore provides a direct time-domain picture of the conversion of the initially static field configuration into an outgoing electromagnetic transient.

\section{Conclusion}
In this work, we investigated static-to-dynamic electromagnetic field conversion in a temporally switched Lorentz medium initially in electrostatic equilibrium. An abrupt change of the oscillator strength leaves the material polarization out of equilibrium with the post-switch medium, providing the initial excitation for the generated transient. Using a Laplace-domain formulation, we derived an analytical residue representation of the field and showed that the finite dispersive slab supports both accumulating and remaining poles, with the accumulating families approaching the complex frequencies associated with the Lorentz resonance. The residue expansion was shown to converge numerically and was validated against time-domain COMSOL simulations.

Material memory governs the earliest stage of the response, producing a quadratic field onset and setting the velocity of the earliest disturbance through the high-frequency permittivity. The temporal switch also generates broadband spectral content spanning positive-, near-zero-, and negative-permittivity regimes, which exhibit distinct spatial field distributions. Their superposition gives rise to the causal outgoing transient in air. These results demonstrate how material dispersion and polarization memory shape the generation and evolution of electromagnetic transients produced from an initially static field.

\bibliography{sample}

\begin{thebibliography}{10}
\newcommand{\enquote}[1]{``#1''}

\bibitem{Morgenthaler58}
F.~R. Morgenthaler, \enquote{{Velocity modulation of electromagnetic waves},}
  {\protect\JournalTitle{IEEE Transactions on Microwave Theory and Techniques}}
  \textbf{6}, 167--172 (1958).

\bibitem{Felsen70}
L.~B. Felsen and G.~M. Whitman, \enquote{{Wave propagation in time-varying
  media},} {\protect\JournalTitle{IEEE Transactions on Antennas and
  Propagation}} \textbf{18}, 242--253 (1970).

\bibitem{Fante71}
R.~L. Fante, \enquote{{Transmission of electromagnetic waves into time-varying
  media},} {\protect\JournalTitle{IEEE Transactions on Antennas and
  Propagation}} \textbf{19}, 417--424 (1971).

\bibitem{Oliner61}
A.~A. Oliner and A.~Hessel, \enquote{{Wave propagation in a medium with a
  progressive sinusoidal disturbance},} {\protect\JournalTitle{IEEE
  Transactions on Microwave Theory and Techniques}} \textbf{9}, 337--343
  (1961).

\bibitem{Salary18}
M.~M. Salary, S.~Jafar-Zanjani, and H.~Mosallaei, \enquote{{Time-varying
  metamaterials based on graphene-wrapped microwires: Modeling and potential
  applications},} {\protect\JournalTitle{Physical Review B}} \textbf{97},
  115421 (2018).

\bibitem{Wang23}
X.~Wang, M.~S. Mirmoosa, V.~S. Asadchy, \emph{et~al.},
  \enquote{{Metasurface-based realization of photonic time crystals},}
  {\protect\JournalTitle{Science Advances}} \textbf{9}, eadg7541 (2023).

\bibitem{Zhou20}
Y.~Zhou, M.~Z. Alam, M.~Karimi, \emph{et~al.}, \enquote{{Broadband frequency
  translation through time refraction in an epsilon-near-zero material},}
  {\protect\JournalTitle{Nature Communications}} \textbf{11}, 2180 (2020).

\bibitem{Ptitcyn23}
G.~Ptitcyn, M.~S. Mirmoosa, S.~Hrabar, and S.~A. Tretyakov,
  \enquote{{Time-modulated circuits and metasurfaces for emulating arbitrary
  transfer functions},} {\protect\JournalTitle{Physical Review Applied}}
  \textbf{20}, 014041 (2023).

\bibitem{Engheta21}
N.~Engheta, \enquote{{Metamaterials with high degrees of freedom: Space, time,
  and more},} {\protect\JournalTitle{Nanophotonics}} \textbf{10}, 639--642
  (2021).

\bibitem{Engheta23}
N.~Engheta, \enquote{{Four-dimensional optics using time-varying
  metamaterials},} {\protect\JournalTitle{Science}} \textbf{379}, 1190--1191
  (2023).

\bibitem{Solis21}
D.~M. Sol{\'i}s and N.~Engheta, \enquote{{Functional analysis of the
  polarization response in linear time-varying media: A generalization of the
  Kramers--Kronig relations},} {\protect\JournalTitle{Physical Review B}}
  \textbf{103}, 144303 (2021).

\bibitem{Solis21b}
D.~M. Sol{\'i}s, R.~Kastner, and N.~Engheta, \enquote{{Time-varying materials
  in the presence of dispersion: Plane-wave propagation in a Lorentzian medium
  with temporal discontinuity},} {\protect\JournalTitle{Photonics Research}}
  \textbf{9}, 1842--1853 (2021).

\bibitem{Koutserimpas24}
T.~T. Koutserimpas and F.~Monticone, \enquote{{Time-varying media, dispersion,
  and the principle of causality},} {\protect\JournalTitle{Optical Materials
  Express}} \textbf{14}, 1222--1236 (2024).

\bibitem{VazquezLozano23}
J.~E. V{\'a}zquez-Lozano and I.~Liberal, \enquote{{Shaping the quantum vacuum
  with anisotropic temporal boundaries},}
  {\protect\JournalTitle{Nanophotonics}} \textbf{12}, 539--548 (2023).

\bibitem{Mirmoosa24}
M.~S. Mirmoosa, M.~H. Mostafa, A.~Norrman, and S.~A. Tretyakov, \enquote{{Time
  interfaces in bianisotropic media},} {\protect\JournalTitle{Physical Review
  Research}} \textbf{6}, 013334 (2024).

\bibitem{Vezzoli18}
S.~Vezzoli, V.~Bruno, C.~DeVault, \emph{et~al.}, \enquote{{Optical time
  reversal from time-dependent epsilon-near-zero media},}
  {\protect\JournalTitle{Physical Review Letters}} \textbf{120}, 043902 (2018).

\bibitem{Moussa23}
H.~Moussa, G.~Xu, S.~Yin, \emph{et~al.}, \enquote{{Observation of temporal
  reflections and broadband frequency translations at photonic time
  interfaces},} {\protect\JournalTitle{Nature Physics}} \textbf{19}, 863--868
  (2023).

\bibitem{Lustig23}
E.~Lustig, O.~Segal, S.~Saha, \emph{et~al.}, \enquote{{Time-refraction optics
  with single-cycle modulation},} {\protect\JournalTitle{Nanophotonics}}
  \textbf{12}, 2221--2230 (2023).

\bibitem{Wu20}
Z.~Wu and A.~Grbic, \enquote{{Serrodyne frequency translation using
  time-modulated metasurfaces},} {\protect\JournalTitle{IEEE Transactions on
  Antennas and Propagation}} \textbf{68}, 1599--1606 (2020).

\bibitem{Wu20b}
Z.~Wu, C.~Scarborough, and A.~Grbic, \enquote{{Space--time-modulated
  metasurfaces with spatial discretization: Free-space $N$-path systems},}
  {\protect\JournalTitle{Physical Review Applied}} \textbf{14}, 064060 (2020).

\bibitem{Lee18}
K.~Lee, J.~Son, J.~Park, \emph{et~al.}, \enquote{{Linear frequency conversion
  via sudden merging of meta-atoms in time-variant metasurfaces},}
  {\protect\JournalTitle{Nature Photonics}} \textbf{12}, 765--773 (2018).

\bibitem{Koutserimpas18}
T.~T. Koutserimpas and R.~Fleury, \enquote{{Nonreciprocal gain in non-Hermitian
  time-Floquet systems},} {\protect\JournalTitle{Physical Review Letters}}
  \textbf{120}, 087401 (2018).

\bibitem{Wang18}
N.~Wang, Z.-Q. Zhang, and C.~T. Chan, \enquote{{Photonic Floquet media with a
  complex time-periodic permittivity},} {\protect\JournalTitle{Physical Review
  B}} \textbf{98}, 085142 (2018).

\bibitem{Qin14}
S.~Qin, Q.~Xu, and Y.~E. Wang, \enquote{{Nonreciprocal components with
  distributedly modulated capacitors},} {\protect\JournalTitle{IEEE
  Transactions on Microwave Theory and Techniques}} \textbf{62}, 2260--2272
  (2014).

\bibitem{Sounas14}
D.~L. Sounas and A.~Al{\`u}, \enquote{{Angular-momentum-biased nanorings to
  realize magnetic-free integrated optical isolation},}
  {\protect\JournalTitle{ACS Photonics}} \textbf{1}, 198--204 (2014).

\bibitem{Shi17}
Y.~Shi, S.~Han, and S.~Fan, \enquote{{Optical circulation and isolation based
  on indirect photonic transitions of guided resonance modes},}
  {\protect\JournalTitle{ACS Photonics}} \textbf{4}, 1639--1645 (2017).

\bibitem{Chamanara17}
N.~Chamanara, S.~Taravati, Z.-L. Deck-L{\'e}ger, and C.~Caloz,
  \enquote{{Optical isolation based on space--time engineered asymmetric
  photonic band gaps},} {\protect\JournalTitle{Physical Review B}} \textbf{96},
  155409 (2017).

\bibitem{Dinc17}
T.~Dinc, M.~Tymchenko, A.~Nagulu, \emph{et~al.}, \enquote{{Synchronized
  conductivity modulation to realize broadband lossless magnetic-free
  non-reciprocity},} {\protect\JournalTitle{Nature Communications}} \textbf{8},
  795 (2017).

\bibitem{Fleury18}
R.~Fleury, D.~L. Sounas, and A.~Al{\`u}, \enquote{{Non-reciprocal optical
  mirrors based on spatio-temporal acousto-optic modulation},}
  {\protect\JournalTitle{Journal of Optics}} \textbf{20}, 034007 (2018).

\bibitem{Taravati20}
S.~Taravati and G.~V. Eleftheriades, \enquote{{Full-duplex nonreciprocal beam
  steering by time-modulated phase-gradient metasurfaces},}
  {\protect\JournalTitle{Physical Review Applied}} \textbf{14}, 014027 (2020).

\bibitem{Fang12}
K.~Fang, Z.~Yu, and S.~Fan, \enquote{{Realizing effective magnetic field for
  photons by controlling the phase of dynamic modulation},}
  {\protect\JournalTitle{Nature Photonics}} \textbf{6}, 782--787 (2012).

\bibitem{Fang14}
K.~Fang, Z.~Yu, and S.~Fan, \enquote{{Experimental demonstration of a photonic
  Aharonov--Bohm effect at radio frequencies},} {\protect\JournalTitle{Physical
  Review Letters}} \textbf{112}, 053901 (2014).

\bibitem{Mirmoosa19}
M.~S. Mirmoosa, G.~A. Ptitcyn, V.~S. Asadchy, and S.~A. Tretyakov,
  \enquote{{Time-varying reactive elements for extreme accumulation of
  electromagnetic energy},} {\protect\JournalTitle{Physical Review Applied}}
  \textbf{11}, 014024 (2019).

\bibitem{Hecht23}
K.~Hecht, D.~Gonz{\'a}lez-Ovejero, D.~L. Sounas, and M.~J. Mencagli,
  \enquote{{First-principles analysis of energy exchange in time-varying
  capacitors for energy trapping applications},} {\protect\JournalTitle{IEEE
  Access}} \textbf{11}, 71494--71502 (2023).

\bibitem{Rizza22}
C.~Rizza, G.~Castaldi, and V.~Galdi, \enquote{{Short-pulsed metamaterials},}
  {\protect\JournalTitle{Physical Review Letters}} \textbf{128}, 257402 (2022).

\bibitem{Shlivinski18}
A.~Shlivinski and Y.~Hadad, \enquote{{Beyond the Bode--Fano bound: Wideband
  impedance matching for short pulses using temporal switching of
  transmission-line parameters},} {\protect\JournalTitle{Physical Review
  Letters}} \textbf{121}, 204301 (2018).

\bibitem{Li19}
H.~Li, A.~Mekawy, and A.~Al{\`u}, \enquote{{Beyond Chu's limit with Floquet
  impedance matching},} {\protect\JournalTitle{Physical Review Letters}}
  \textbf{123}, 164102 (2019).

\bibitem{Yang22}
X.~Yang, E.~Wen, and D.~F. Sievenpiper, \enquote{{Broadband time-modulated
  absorber beyond the Bode--Fano limit for short pulses by energy trapping},}
  {\protect\JournalTitle{Physical Review Applied}} \textbf{17}, 044003 (2022).

\bibitem{Fritts25}
Z.~Fritts, A.~Babaee, S.~M. Young, and A.~Grbic, \enquote{{Space--time
  modulation of a multimode electrically small antenna for increased matching
  and efficiency bandwidths},} {\protect\JournalTitle{IEEE Transactions on
  Antennas and Propagation}} \textbf{73}, 1308--1320 (2025).

\bibitem{OlguinLopez26}
M.~Olguin-Lopez, D.~L. Sounas, N.~Engheta, \emph{et~al.}, \enquote{{Aperiodic
  temporal modulation for distortionless broadband impedance matching beyond
  the Bode--Fano limit},} {\protect\JournalTitle{arXiv preprint
  arXiv:2608.14933}}  (2026).

\bibitem{Firestein23}
C.~Firestein, A.~Shlivinski, and Y.~Hadad, \enquote{{Sum rule bounds beyond
  Rozanov criterion in linear and time-invariant thin absorbers},}
  {\protect\JournalTitle{Physical Review B}} \textbf{108}, 014308 (2023).

\bibitem{Hayran24}
Z.~Hayran and F.~Monticone, \enquote{{Beyond the Rozanov bound on
  electromagnetic absorption via periodic temporal modulations},}
  {\protect\JournalTitle{Physical Review Applied}} \textbf{21}, 044007 (2024).

\bibitem{Ginzburg82}
V.~L. Ginzburg, \enquote{{Transition radiation and transition scattering},}
  {\protect\JournalTitle{Physica Scripta}} \textbf{T2A}, 182--191 (1982).

\bibitem{Wilks89}
S.~C. Wilks, J.~M. Dawson, W.~B. Mori, \emph{et~al.}, \enquote{{Photon
  accelerator},} {\protect\JournalTitle{Physical Review Letters}} \textbf{62},
  2600--2603 (1989).

\bibitem{Mori95}
W.~B. Mori, T.~Katsouleas, J.~M. Dawson, and C.~H. Lai, \enquote{{Conversion of
  dc fields in a capacitor array to radiation by a relativistic ionization
  front},} {\protect\JournalTitle{Physical Review Letters}} \textbf{74},
  542--545 (1995).

\bibitem{Lai96}
C.~H. Lai, R.~Liou, T.~C. Katsouleas, \emph{et~al.}, \enquote{{Demonstration of
  microwave generation from a static field by a relativistic ionization front
  in a capacitor array},} {\protect\JournalTitle{Physical Review Letters}}
  \textbf{77}, 4764--4767 (1996).

\bibitem{Esarey96}
E.~Esarey, P.~Sprangle, B.~Hafizi, and P.~Serafim, \enquote{{Radiation
  generation by photoswitched, periodically biased semiconductors},}
  {\protect\JournalTitle{Physical Review E}} \textbf{53}, 6419--6426 (1996).

\bibitem{Murphy06}
C.~D. Murphy, R.~Trines, J.~Vieira, \emph{et~al.}, \enquote{{Evidence of photon
  acceleration by laser wake fields},} {\protect\JournalTitle{Physics of
  Plasmas}} \textbf{13}, 033108 (2006).

\bibitem{Mencagli2022}
M.~J. Mencagli, D.~L. Sounas, M.~Fink, and N.~Engheta,
  \enquote{Static-to-dynamic field conversion with time-varying media,}
  {\protect\JournalTitle{Phys. Rev. B}} \textbf{105}, 144301 (2022).

\bibitem{Spiegel1986}
M.~R. Spiegel, \emph{Laplace Transforms} (McGraw-Hill Education, New York,
  1986).

\bibitem{supp}
\enquote{See supplemental material at,} \url{http://........} For additional
  analytical details, including the derivation of the Laplace-domain field
  equations, the asymptotic characterization of the accumulating pole families,
  the contour deformation used for the Bromwich inversion, and the derivation
  of the residue representation.

\end{thebibliography}


\begin{thebibliography}{1}

\bibitem[S1]{Spiegel1986SM}
M.~R. Spiegel,
\textit{Laplace Transforms}
(McGraw-Hill Education, New York, 1986).

\end{thebibliography}



\clearpage

\begingroup
\centering
{\large\bfseries Supplemental Material\par}
\vspace{6pt}
{\large Static-to-dynamic field conversion in a temporally switched Lorentz medium\par}
\endgroup

\vspace{1em}

\setcounter{section}{0}
\setcounter{subsection}{0}
\setcounter{equation}{0}
\setcounter{figure}{0}
\setcounter{table}{0}

\renewcommand{\thesection}{S\arabic{section}}
\renewcommand{\thesubsection}{\thesection.\arabic{subsection}}
\renewcommand{\theequation}{S\arabic{equation}}
\renewcommand{\thefigure}{S\arabic{figure}}
\renewcommand{\thetable}{S\arabic{table}}

\bigskip

\section{Derivation of the wave equations and Laplace-domain field solutions}
\label{sec:wave-equation-derivation}
In this section, we provide the detailed derivation of the wave equations and the corresponding Laplace-domain electric-field solutions stated in Sec.~3 of the main text.

For the TEM waves considered in this work, the dynamic electric and magnetic fields in the Lorentz medium can be written as
\begin{align}
    \mathbf e_{\LL}^\dd(z,t)
    &=\hat{\mathbf x}\,e_{\LL}^\dd(z,t),\\
     \mathbf h_{\LL}^\dd(z,t)
    &= \hat{\mathbf y}\,h_{\LL}^\dd(z,t).
\end{align}
These fields satisfy Maxwell's equations
\begin{align}
    \frac{\partial e_{\LL}^\dd}{\partial z}
    +\mu_0\frac{\partial h_{\LL}^\dd}{\partial t}
    &=0,
    \label{faraday-time}\\
    \frac{\partial h_{\LL}^\dd}{\partial z}
    +\varepsilon_0\varepsilon_\infty
    \frac{\partial e_{\LL}^\dd}{\partial t}
    +\frac{\partial p_{\LL}^\dd}{\partial t}
    &=0.
    \label{ampere-time}
\end{align}
The dynamic fields are subject to the initial conditions $e_{\LL}^\dd(z,0^+)=0$ and $h_{\LL}^\dd(z,0^+)=0$, which follow from the continuity of the total electromagnetic fields across the temporal interface and from the absence of a magnetic field in the initial static state. Applying the unilateral Laplace transform \cite{Spiegel1986SM},
\begin{equation}
    F(z,s)=\int_0^\infty f(z,t)e^{-st}\,dt,
\end{equation}
to Eqs.~\eqref{faraday-time} and \eqref{ampere-time} and using initial conditons for the electric and magnetic fields defined above, we obtain
\begin{align}
    \frac{\partial} {\partial z}E_{L}^{\mathrm d}(z, s)
    +s\mu_0H_{L}^{\mathrm d}(z, s)&=0,
    \label{faraday-laplace}\\
    \frac{\partial}{\partial z}H_{L}^{\mathrm d}(z, s)
    +s\varepsilon_0\varepsilon_\infty E_{L}^{\mathrm d}(z, s)
    +s P_{\LL}^\dd - p_{\LL}^\dd(z,0^+)
    &=0.
    \label{ampere-laplace-P}
\end{align}
Using the Lorentz polarization relation introduced in the main text [Eq.~(7)]
\begin{equation}
    P_{\LL}^\dd(z,s)
    =\varepsilon_0
    \frac{\omega_{p2}^2}{A_0(s)}
    E_{\LL}^\dd(z, s)
    +\frac{s+\gamma}{A_0(s)}\Delta P
\end{equation}
together with the polarization initial condition [Eq.~5]
\begin{equation}
    p_{\LL}^\dd(z,0^+)= \Delta P
    = \varepsilon_0 \frac{\omega_{p1}^2-\omega_{p2}^2}{\omega_0^2} E^\s,\label{dynamic_pol_initial_cond1}
\end{equation}
Eq.~\eqref{ampere-laplace-P} becomes
\begin{equation}
    \frac{\partial}{\partial z}H_{\LL}^\dd(z, s)
    +s\varepsilon_0\left(\varepsilon_\infty+\frac{\omega_{p2}^2}{A_0(s)}\right)
    E_{\LL}^\dd(z, s)
    =\Delta P \frac{\omega_0^2}{A_0(s)}.
    \label{ampere-laplace}
\end{equation}
Combining Eqs.~(\ref{ampere-laplace}) and (\ref{faraday-laplace}) to eliminate $H_L^{\mathrm d}(z,s)$ leads to the inhomogeneous wave equation
\begin{equation}
    \frac{\partial^2}{\partial z^2}E_{\LL}^\dd(z,s)
    -
    k_{\LL}^2(s)E_{\LL}^\dd(z,s)
    =
    -s\mu_0\Delta P\frac{\omega_0^2}{A_0(s)},
    \label{wave_equation1}
\end{equation}
where
\begin{equation}
k_L^2(s)=\frac{s^2}{c_a^2}\varepsilon_2(s),
\qquad
\varepsilon_2(s)=\varepsilon_\infty\frac{A_1(s)}{A_0(s)},
\label{kd-final}
\end{equation}
with $k_\mathrm{a} = \frac{s}{c_\mathrm{a}}$, $c_\mathrm{a} = \frac{1}{\sqrt{\mu_0\varepsilon_0}}$, $A_1(s) = s^2+\gamma s+\omega_1^2$, and $\omega_1^2 = \omega_0^2+\frac{\omega_{p2}^2}{\varepsilon_\infty}$.
In the air region, where no material polarization is present, the corresponding dynamic electric field satisfies the homogeneous equation
\begin{equation}
    \frac{\partial^2{E}_{\mathrm a}^{\dd}}{\partial z^2}
    -k_\mathrm{a}^2{E}_{\mathrm a}^{\dd}=0.
    \label{air-wave}
\end{equation}
Solving the wave equations in the Lorentz and air regions and imposing the spatial boundary conditions, which include field symmetry about $z=0$, continuity of the tangential electric and magnetic fields at $z=L$, and the outgoing-wave condition in the air region, leads to the following Laplace-domain electric fields:
\begin{equation}\label{E_field_lorentz_L}
    E_{\LL}^\dd(z, s) = E_{\mathrm{part}} \left[ 1 - \frac{k_{\mathrm a} \cosh(k_L z)}{k_{\mathrm a} \cosh(k_{\LL} L) + k_{\LL} \sinh(k_{\LL} L)} \right],
    \qquad 0\leq z\leq L,
\end{equation}
and
\begin{equation}\label{E_field_air_L}
    E_{\mathrm a}^{\dd}(z,s) = E_{\mathrm{part}} \frac{k_{\LL} \sinh(k_{\LL} L)}{k_\mathrm{a} \cosh(k_{\LL} L) + k_{\LL} \sinh(k_{\LL} L)} e^{-k_\mathrm{a}(z-L)},
    \qquad z>L,
\end{equation}
where
\begin{equation}\label{part_sol}
  E_{\mathrm{part}} = E^\s \frac{\omega_{p1}^2 - \omega_{p2}^2}{s \varepsilon_\infty A_1(s)}=E^\s \left( \frac{\varepsilon_{\s 1}}{\varepsilon_{\s 2}}-1\right)\frac{\omega_1^2}{s A_1(s)}.
\end{equation}

\section{Contour evaluation of the inverse Laplace transform in the Lorentz medium}
\label{sec:inverse_L_transform}
\subsection{Bromwich integral and contour deformation}

The dynamic electric field inside the Lorentz medium is obtained from the Bromwich inversion formula,
\begin{equation}
\mathrm{e}_{\LL}^\dd(z,t) = \frac{E^\s}{2\pi i} \left( \frac{\varepsilon_{\s 1}}{\varepsilon_{\s 2}}-1\right)
    \int_{\sigma_\B-i\infty}^{\sigma_\B+i\infty}
    \Phi(s,z,t)\,ds,
        \qquad t>0,
    \label{eq:SM_Bromwich_integral}
\end{equation}
where the vertical line $\operatorname{Re}(s)=\sigma_\B$ lies to the right of all singularities of the integrand. The Bromwich integrand $\Phi(s,z,t)$ is defined in Eq.~(16) of the main text. For completeness and to facilitate the discussion, we restate it here as
\begin{equation}
    \Phi(s,z,t)
    =
    \frac{1}{sA_1(s)}
    \frac{N(s,z)}{D(s)}
    e^{st},
\end{equation}
with
\begin{equation}
D(s) = \cosh\!\left[k_{\LL}(s)L\right] +\frac{k_{\LL}(s)}{k_a(s)}\sinh\!\left[k_{\LL}(s)L\right]
\label{D_def}
\end{equation}
and
\begin{align}
N(s,z) = D(s)- \cosh\!\left[k_{\LL}(s)z\right].
\label{N_def}
\end{align}
As discussed in the main text, the square-root dependence of $k_\LL(s)$ produces no branch-cut contribution, while $s=0$ and the roots of $A_1(s)$ are removable singularities. The contour analysis can therefore be restricted to the isolated zeros of $D(s)$ and to their accumulation near the roots of $A_0(s)$.

\begin{figure}[t!]
    \centering
    \includegraphics[scale=0.6]{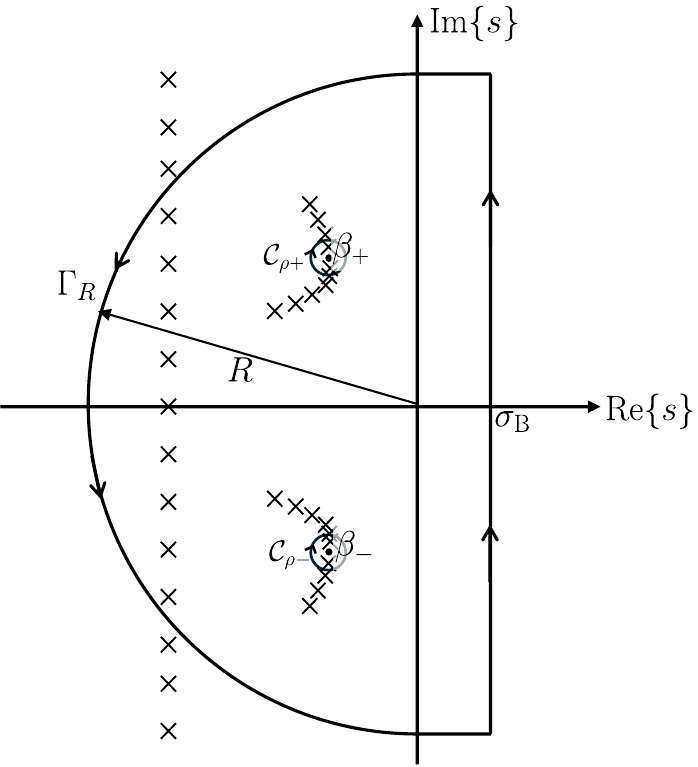}
    \caption{\label{Bromwich_contour}The Bromwich line is closed in the left half of the complex $s$ plane. Neighborhoods of the nonisolated accumulation points $\beta_\pm$ are excluded by clockwise-oriented, pole-avoiding contours. The punctured region contains only finitely many isolated slab-mode poles for finite values of the outer-contour size and the excluded-neighborhood radii.}
\end{figure}

For $t>0$, the Bromwich contour is closed in the left half of the complex $s$ plane [Fig.~\ref{Bromwich_contour}]. Because infinitely many poles accumulate toward the roots of $A_0(s)$, which we denote by $\beta_\pm$, these points cannot be treated as ordinary isolated singularities. We therefore exclude neighborhoods of $\beta_\pm$ from the interior of the closed contour. More specifically, let $\mathcal{C}$ denote a contour consisting of the truncated Bromwich line, a closing contour $\Gamma_R$ in the left half-plane, and two clockwise-oriented contours $\mathcal{C}_{\rho_\pm}$ surrounding the accumulation points. 
The radii $\rho_\pm$ are chosen so that the corresponding contours do not intersect or approach any of the isolated poles accumulating near $\beta_\pm$.

For finite $R$ and nonzero $\rho_\pm$, the punctured region enclosed by $\mathcal{C}$ contains only finitely many isolated poles of $\Phi(s,z,t)$. The residue theorem therefore gives
\begin{align}
    \frac{1}{2\pi i}
    \int_{\sigma_\mathrm{B}-iR}^{\sigma_\mathrm{B}+iR}
        \Phi(s,z,t)\,ds
    &=
    \sum_{s_n\in\mathcal{P}_{R,\rho}}
        \operatorname{Res}_{s=s_n}\Phi(s,z,t)
    \nonumber\\
    &\quad
    -
    \frac{1}{2\pi i}
    \int_{\Gamma_R}
        \Phi(s,z,t)\,ds
    \nonumber\\
    &\quad
    -
    \frac{1}{2\pi i}
    \sum_{\nu\in\{+,-\}}
    \int_{\mathcal{C}_{\rho_\nu}}
        \Phi(s,z,t)\,ds .
    \label{eq:SM_truncated_Bromwich}
\end{align}
where $\mathcal{P}_{R,\rho}$ denotes the set of isolated poles enclosed by the punctured contour.
This requires establishing that the contribution from the closing contour $\Gamma_R$ tends to zero as $R\to\infty$, and evaluating the contributions from the contours surrounding $\beta_\pm$ as $\rho_\pm\to0$ along admissible pole-avoiding sequences. Before evaluating these contour contributions, we characterize the local distribution of the dispersion-induced slab-mode poles near the accumulation points.
\subsection{Pole accumulation near the material resonances}
We now characterize the distribution of the slab-mode poles near the roots of $A_0(s)$. We denote these roots by
\begin{equation}
    \beta_\pm = -\frac{\gamma}{2} \pm i\sqrt{ \omega_0^2-\frac{\gamma^2}{4}}.
    \label{eq:SM_beta_pm}
\end{equation}
Note that the damped regime, $2\omega_0>\gamma$, is assumed, so that $\beta_\pm$ form a complex-conjugate pair.
Let $\beta$ denote either root of $A_0(s)$. Using $A_0(s) = A_0'(\beta)(s-\beta) + O\!\left[(s-\beta)^2\right]$, with $A_0'(\beta)=2\beta+\gamma$, and Eq.~(\ref{kd-final}), we obtain
\begin{equation}
    \left[k_{\LL}(s)L\right]^2 = \frac{U_\beta}{s-\beta}+O(1),
    \qquad
    s \to \beta,
    \label{eq:SM_kL_local}
\end{equation}
where $U_\beta = \frac{ \varepsilon_\infty L^2\beta^2 A_{1}(\beta)}{ c_{\mathrm a}^2 A_0'(\beta)}$. Eq.~(\ref{eq:SM_kL_local}) can be asymptotically inverted to give
\begin{equation}
    s-\beta = \frac{U_\beta}{[k_{\LL}(s)L]^2}+O\!\left([k_{\LL}(s)L]^{-4}\right).
    \label{eq:SM_s_kL_local}
\end{equation}
The slab-mode poles are associated with the zeros of $D(s)$, defined in Eq.~(\ref{D_def}). As $s\to\beta$, one has $|k_{\LL}(s)L|\to\infty$. The large-order solutions of $D(s)=0$ approach the zeros of the hyperbolic sine and satisfy
\begin{equation}
    k_{\LL}\!\left(s_m^{(\beta)}\right)L = i\pi m +O\!\left(m^{-1}\right),
    \qquad
    m\rightarrow\infty .
    \label{eq:SM_kL_pole_asymptotic}
\end{equation}
Substitution of Eq.~(\ref{eq:SM_kL_pole_asymptotic}) into Eq.~(\ref{eq:SM_s_kL_local}) gives
\begin{equation}
    s_m^{(\beta)} = \beta -
    \frac{U_\beta}{\pi^2m^2}+O\!\left(m^{-4}\right),
    \qquad
    m\rightarrow\infty .
    \label{eq:SM_pole_accumulation}
\end{equation}


\noindent This equation shows that infinitely many isolated zeros of \(D(s)\) accumulate toward each root of \(A_0(s)\). Their asymptotic locations depend on both the slab length and the Lorentz-dispersion parameters, as seen from the definition of \(U_\beta\). In the following, we refer to these zeros, and to the corresponding poles of the Bromwich integrand, as accumulating poles.

In addition to the families of accumulating poles associated with \(\beta_\pm\), \(D(s)\) may possess other isolated zeros that remain separated from the roots of \(A_0(s)\). We refer to these as remaining poles. These require no special treatment in the contour deformation, and together with the accumulating poles they constitute the full set of isolated poles entering the residue representation.

At a zero $s_n$ of $D(s)$, the numerator reduces to $N(s_n,z)=-\cosh[k_{\LL}(s_n)z]$. Hence, the zero of $D(s)$ is not canceled identically throughout the Lorentz medium. At a fixed interior observation point, however, the corresponding residue may vanish if $\cosh[k_L(s_n)z]=0$. Such a vanishing residue would represent a node of that modal contribution at the selected position, rather than the absence of the mode from the complete spatial field. 

\subsection{Contribution from the large-$|s|$ contour}
We next examine the contribution from the closing contour $\Gamma_R$ (Fig.~\ref{Bromwich_contour}) as $R\rightarrow\infty$. The contours $\Gamma_R$ are selected from an admissible sequence that does not intersect, or pass arbitrarily close to, any isolated zero of $D(s)$.

For large $|s|$, the material factors satisfy
\begin{equation}
    \frac{A_1(s)}{A_0(s)} = 1+ \frac{\omega_1^2-\omega_0^2}{s^2} + O\!\left(|s|^{-3}\right),
\end{equation}
and hence
\begin{align}
    \frac{k_{\LL}(s)}{k_{\mathrm a}(s)}
    &=\sqrt{\varepsilon_\infty} + O\!\left(|s|^{-2}\right),\label{SM_large_s_kLka}\\
     k_{\LL}(s)
    &= \frac{\sqrt{\varepsilon_\infty}}{c_{\mathrm a}}s+
    O\!\left(|s|^{-1}\right).\label{SM_large_s_kL}
\end{align}
The slab denominator [Eq.~(\ref{D_def})] may be expressed as
\begin{equation}
    D(s) = \frac{1}{2}
    \left[
        \left(
            1+\frac{k_{\LL}(s)}{k_{\mathrm a}(s)}
        \right)e^{k_{\LL}(s)L}
        +
        \left(
            1-\frac{k_{\LL}(s)}{k_{\mathrm a}(s)}
        \right)e^{-k_{\LL}(s)L}
    \right].
    \label{eq:SM_large_s_D}
\end{equation}
Because the complete integrand is invariant under $k_L(s)\mapsto-k_L(s)$, the sign of the square root may be chosen conveniently when estimating its magnitude. On the closing contour, we take $\operatorname{Re}[k_L(s)]\leq0$. 

For $\varepsilon_\infty\neq1$, the coefficient multiplying $e^{-k_L(s)L}$ in Eq.~(\ref{eq:SM_large_s_D}) approaches the nonzero constant $1-\sqrt{\varepsilon_\infty}$. Away from the zeros of $D(s)$, this gives
\begin{equation}
    \frac{\cosh[k_L(s)z]}{D(s)}
    =
    O\!\left(
        e^{\operatorname{Re}[k_L(s)](L-z)}
    \right)
    =
    O(1),
    \qquad 0\leq z\leq L .
    \label{eq:SM_large_s_ratio}
\end{equation}
The pole-avoiding choice of $\Gamma_R$ ensures that this bound remains uniform along the admissible sequence of closing contours. From the previous equation, we obtain 
\begin{equation}
    \frac{N(s,z)}{D(s)} = O(1),
    \qquad |s|\rightarrow\infty .
\end{equation}
Moreover,
\begin{equation}
    \frac{\omega_L^2}{sA_L(s)}
    =
    O\!\left(|s|^{-3}\right).
\end{equation}
Consequently, the complete Bromwich integrand satisfies
\begin{equation}
    \Phi(s,z,t) = O\!\left(|s|^{-3}e^{\operatorname{Re}(s)t}\right), \qquad |s|\rightarrow\infty
\end{equation}
along $\Gamma_R$.

Since $\operatorname{Re}(s)\leq\sigma_{\B}$ on the closing contour,
\begin{equation}
    |e^{st}|
    =
    e^{t\operatorname{Re}(s)}
    \leq
    e^{\sigma_{\B}t},
\end{equation}
while the length of $\Gamma_R$ scales as $O(R)$. It follows that
\begin{align}
    \left|
        \int_{\Gamma_R}
        \Phi(s,z,t)\,ds
    \right|
    &\leq
    \operatorname{length}(\Gamma_R)
    \max_{s\in\Gamma_R}
    |\Phi(s,z,t)|
    \nonumber\\
    &=
    O(R)\,O(R^{-3})
    =
    O(R^{-2}).
\end{align}
Therefore,
\begin{equation}
    \lim_{R\rightarrow\infty}
    \int_{\Gamma_R}
    \Phi(s,z,t)\,ds
    =
    0,
    \qquad t>0,
    \label{eq:SM_large_contour_vanishes}
\end{equation}
along the admissible pole-avoiding sequence. Thus, the large-$|s|$ closing contour provides no additional contribution to the inverse Laplace transform.

When $\varepsilon_\infty=1$, the first term in the large-$|s|$ expansion of $D(s)$ vanishes, and the second-order term must be retained. However, following an analysis similar to that for $\varepsilon_\infty\neq1$, it can be shown that the contribution from the large-$|s|$ closing contour still vanishes along an admissible pole-avoiding sequence.

\subsection{Contributions from the accumulation-point neighborhoods}
We next evaluate the contributions from the contours surrounding the roots $\beta_\pm$ of $A_0(s)$ (Fig.~\ref{Bromwich_contour}). Because infinitely many isolated poles accumulate toward each of these points, arbitrary circular contours in the complex $s$ plane are not suitable: as their radii decrease, they may intersect or pass arbitrarily close to zeros associated with the accumulating pole families. We therefore consider admissible sequences of pole-avoiding contours.

Differentiating Eq.~(\ref{eq:SM_s_kL_local}) gives
\begin{equation}
    ds = \left\{-2
        \frac{U_\beta}{[k_{\LL}(s)L]^3}
        +
        O\!\left([k_{\LL}(s)L]^{-5}\right)
        \right\} d[k_{\LL}(s)L].
    \label{eq:SM_ds_accumulation}
\end{equation}
We construct a sequence of contours $\mathcal{C}_j^{(\beta)}$ in the $s$-plane as the level curves
\begin{equation}
    |k_{\LL}(s)L|=R_j, \qquad R_j\rightarrow\infty,
\end{equation}
with the values $R_j$ chosen so that the contours remain separated from the zeros of $D(s)$. From Eq.~(\ref{eq:SM_s_kL_local}), these contours shrink toward $\beta$ with characteristic radius $\rho_j=O(R_j^{-2})$. Along this admissible sequence, the ratio $N(s,z)/D(s)$ remains bounded because the contours are separated from the zeros of $D(s)$. Moreover, the factors $1/s$, $1/A_1(s)$, and $e^{st}$ remain finite as $s\rightarrow\beta$, since $\beta\neq0$ and $A_1(\beta)\neq0$. Hence, $\Phi(s,z,t)=O(1)$ on $\mathcal{C}_{j}^{(\beta)}$. Using
Eq.~(\ref{eq:SM_ds_accumulation}), the contour integral therefore
satisfies
\begin{align}
    \left|
        \int_{\mathcal{C}_{j}^{(\beta)}}
        \Phi(s,z,t)\,ds
    \right| = O\!\left(R_j^{-2}\right).
\end{align}
Accordingly,
\begin{equation}
    \lim_{j\rightarrow\infty}
    \int_{\mathcal{C}_{j}^{(\beta)}}
        \Phi(s,z,t)\,ds
    =
    0.
    \label{eq:SM_accumulation_contour_vanishes}
\end{equation}
The same conclusion holds for both $\beta_+$ and $\beta_-$.

Thus, although the roots of $A_0(s)$ are nonisolated singularities and accumulation points of infinitely many dispersion-induced slab-mode poles, the pole-avoiding contours shrinking around them produce no additional contribution to the inverse Laplace transform. Their role is instead accounted for through the residues of the individual isolated poles that accumulate toward them.

\subsection{Residue representation of the time-domain field}
The results of the preceding subsections show that, for $t>0$, neither the large-$|s|$ closing contour nor the pole-avoiding contours shrinking around the accumulation points $\beta_\pm$ contribute to the Bromwich inversion. Combining Eq.~(\ref{eq:SM_truncated_Bromwich}) with Eq.~(\ref{eq:SM_Bromwich_integral}), the time-domain field can be expressed as
\begin{equation}
    {\mathrm e}_{\LL}^{\mathrm d}(z,t) =
    E^\s
\left(
\frac{\varepsilon_{\s 1}}{\varepsilon_{\s 2}}-1
\right)
    \sum_{s_n\in\mathcal{P}}
    \operatorname{Res}_{s=s_n}\Phi(s,z,t),
    \qquad t>0,
    \label{eq:SM_residue_limit}
\end{equation}
where $\mathcal{P}$ denotes the infinite isolated poles of $\Phi(s,z,t)$. The residue sum is understood in the contour-ordered sense, namely as the limit of the finite residue sums enclosed by the admissible pole-avoiding contours.

Let $s_n$ be a simple zero of $D(s)$. Since the remaining factors of
$\Phi(s,z,t)$ are analytic at $s_n$, its residue is
\begin{equation}
    \operatorname{Res}_{s=s_n}\Phi(s,z,t) =
    \frac{\omega_1^2 N(s_n,z)}{
        s_n A_1(s_n)D'(s_n)
    }
    e^{s_nt}.
    \label{eq:SM_residue_simple_pole}
\end{equation}
Defining
\begin{equation}
r_n(z)=
\frac{\omega_1^2 N(s_n,z)}
{s_n A_1(s_n)D'(s_n)}.
\label{eq:SM_modal_coefficient}
\end{equation}
the time-domain field can be written compactly as
\begin{equation}
    {\mathrm e}_{\LL}^{\mathrm d}(z,t)  = E^\s
\left(
\frac{\varepsilon_{\s 1}}{\varepsilon_{\s 2}}-1
\right)
    \sum_{s_n\in\mathcal{P}}
    r_n(z)e^{s_nt},
    \qquad t>0.
    \label{eq:SM_modal_expansion}
\end{equation}
Because the material and geometrical parameters are real, the nonreal poles occur in complex-conjugate pairs and the corresponding residues satisfy the same conjugate symmetry. The residue representation can therefore be written by including the single real pole $s_0$ explicitly and retaining only one member of each nonreal conjugate pair, whose combined contribution is expressed through twice the real part. Denoting by $\mathcal{P}_+$ the poles in the upper half of the complex $s$ plane, Eq.~(\ref{eq:SM_modal_expansion}) can be written equivalently as
\begin{equation}
{\mathrm e}_{\LL}^{\mathrm d}(z,t)
=
E^\s
\left(
\frac{\varepsilon_{\s 1}}{\varepsilon_{\s 2}}-1
\right)
\left[
r_0(z)e^{s_0t}
+
2\operatorname{Re}
\left\{
\sum_{s_{n}\in\mathcal{P}_+}
r_{n}(z)e^{s_nt}
\right\}
\right],
\qquad t>0.
\label{modal_expansion}
\end{equation}
The time-domain response is therefore completely determined by the residues of the isolated slab-mode poles. These include both the accumulating poles, which form families approaching $\beta_\pm$, and the remaining poles, which stay separated from the accumulation points. The points $\beta_\pm$ themselves are nonisolated singularities and do not generate additional residue contributions beyond those already contained in Eq.~(\ref{modal_expansion}).

\begingroup
\renewcommand{\refname}{Supplemental References}

\endgroup


%







\end{document}